\documentclass[intlimits,twoside,a4paper]{article}

\usepackage{amsmath,amssymb}
\usepackage{graphicx}
\usepackage{wrapfig}
\usepackage{cite}
\usepackage[T2A]{fontenc}
\usepackage[cp1251]{inputenc}

\usepackage{color}

\usepackage[eqsecnum]{cmpj2}

\newcommand{\be}{\begin{equation}}
\newcommand{\ee}{\end{equation}}
\newcommand{\bea}{\begin{eqnarray}}
\newcommand{\eea}{\end{eqnarray}}
\newcommand{\non}{\nonumber\\}

\newcommand{\ra}{\raisebox{0.05ex}[0cm][0cm]}
\newcommand{\e}{\includegraphics}

\issue{2012}{15}{3}{33705}

\doinumber{10.5488/CMP.15.33705}

\title[Longitudinal relaxation of mechanically clamped crystals]%
{Longitudinal relaxation of mechanically clamped KH$_2$PO$_4$ type crystals}

\author[R.R. Levitskii, I.R. Zachek, A.S. Vdovych]{R.R. Levitskii\refaddr{label1},
        I.R. Zachek\refaddr{label2}, A.S. Vdovych\refaddr{label1}}
\addresses{
\addr{label1} Institute for Condensed Matter Physics of the National
Academy of Sciences of Ukraine, \\ 1 Svientsitskii Str., 79011 Lviv,
Ukraine \addr{label2} Lviv Polytechnic National University, 12
Bandera Str., 79013 Lviv, Ukraine
}

\authorcopyright{R.R. Levitskii, I.R. Zachek, A.S. Vdovych, 2012}

\date{Received January 27, 2012, in final form July 4, 2012}

\begin{document}

\maketitle

\begin{abstract}
Within the framework of a modified proton ordering model of the KH$_{2}$PO$_{4}$ family ferroelectric crystals, taking into account a linear over the strain $\varepsilon_{6}$ contribution into the proton system energy, we obtain
an expression for longitudinal dynamic dielectric permittivity of a mechanically clamped crystal using the four-particle cluster approximation and the dynamic Glauber approach. At a proper choice of the model parameters, we obtain a good quantitative description of available experimental data for these crystals.
\keywords ferroelectric, cluster approximation, dielectric permittivity, relaxation times
\pacs 77.84.-s, 77.22.-d, 77.80.-e, 77.22.Ch


\end{abstract}

\section{Introduction}
In the late 1960-ies, most theoretical and experimental studies of
ferroelectrics concentrated on various dynamic phenomena. The dispersion of dielectric
permittivity of ferroelectrics was explored  at low frequencies, which provided an important
information on the mechanisms of phase transitions and revealed the peculiarities of the low-frequency dynamics of a  system.
Ferroelectric dispersion is closely related to the presence of a low-frequency excitation, i.e., a soft mode which can be either
resonant or relaxational. Ferroelectric compounds of the  KH$_2$PO$_4$ family occupy an intermediate
position. The region of fundamental dispersion in these crystals is located in the
submillimeter range $\nu \approx~50$~GHz. At deuteration, the ferroelectric dispersion in these crystals
is shifted to the millimeter and microwave ranges.

The major task of dielectric spectra studies of ferroelectric crystals is to explore the
peculiarities of the soft mode behavior, especially in the phase transition region~\cite{145x}. As a rule, the soft  modes
in the  KH$_2$PO$_4$ family ferroelectrics are strongly damped. To explore their character is a complicated task. One has to explore the dielectric spectra of these crystals in a wide frequency range that includes several regions requiring
specific and unique experimental methods of measurements. There is hardly any experimental group
 fully equipped for such studies. This fact, along with the principal
difficulties in experimental measurements of dielectric spectra, and the dependence of  $\hat \varepsilon^*(\omega, T)$ on sample quality and surface treatment,
causes the situation when the experimental data for dielectric spectra of the
KH$_2$PO$_4$ family ferroelectrics turn out to be disembodied and quite conflicting. This should be kept in mind while analysing the  experimental data and
the theoretical results for dynamic characteristics of ferroelectrics including those of the  KH$_2$PO$_4$ family.

In the late 1970-ies, the obtained experimental results for the dynamic characteristics in
the KH$_2$PO$_4$ family compounds  were interpreted mostly within phenomenological
models (see~\cite{145x,58x,60x}).  Phenomenological theories do not make it possible
to reveal the microscopic nature of the dispersion of dielectric
permittivity or to appropriately describe the effect of various factors
on the character of its temperature and frequency dependencies.
The attempts to solve this problem using the Green's function method or Bloch kinetic equations method failed~\cite{16x,17x}.

A vast majority of studies on the theory of relaxation phenomena in
the KH$_2$PO$_4$ family ferroelectrics are based on the stochastic
Glauber model~\cite{61x}. For the first time, the relaxation dynamics
of the KD$_2$PO$_4$ type ferroelectrics was studied using this method
in~\cite{58x}, where, within the four-particle cluster approximation (FPCA), there was initiated a study of the main regularities of longitudinal relaxation in the
case of a paraelectric phase. However, long-range interactions
were not taken into account therein, and the corresponding experimental data for
the KD$_2$PO$_4$ type ferroelectrics were not discussed.
Later on~\cite{342x,343x,344x}, a  more consistent  model of deuterated
 KD$_2$PO$_4$ type ferroelectrics  and ND$_4$D$_2$PO$_4$ type  antiferroelectrics was explored.
 Within the framework of this model, using the FPCA for short-range interactions and the mean field approximation
for long-range interactions, longitudinal dynamic characteristics
of these crystals were calculated. It was shown~\cite{163x,165x,uni2009} that the theory proposed in~\cite{342x,343x,344x} provides a satisfactory
description of thermodynamic and longitudinal dynamic characteristics of the KH$_2$PO$_4$
type ferroelectrics.
In~\cite{48pok,135x,137x}, the authors attempted to develop a more consistent theory of the  KH$_2$PO$_4$ family ferroelectrics in the FPCA which takes  tunneling ($\Omega$)  into account. The results were not good enough to appropriately describe the available experimental data
for the dynamic characteristics of these crystals. However,  the fact of suppression of the
dynamic characteristics of the  KH$_2$PO$_4$ type ferroelectrics by short-range interactions was established. An  effective tunneling parameter
$\tilde \Omega$ ($\tilde \Omega \ll \Omega$) renormalized by the short-range interactions was obtained. It should be noted that the established in~\cite{48pok,135x,137x} suppression of
dynamic characteristics of the  KH$_2$PO$_4$ type ferroelectrics by short-range correlations is the most probable
reason of the Debye-type dispersion of  dielectric permittivity observed in these crystals.

In~\cite{147pok,289pok,120pok},  thermodynamic and dynamic
characteristics of quasi-one-dimensional
hydrogen bonded CsH$_2$PO$_4$ ferroelectrics were found using a self-consistent approach to the calculation
of thermodynamic and dynamic characteristics of pseudospin systems with
essential short-range and long-range interactions, based on the calculation of the free energy functional with
 short-range interactions taken into account in the reference approach.
  It was
established that an essential suppression of the soft vibration mode
by short-range correlations takes place in a wide temperature
range. This fact, just like in the case of KH$_2$PO$_4$, is directly related to the Debye type of longitudinal dielectric permittivity
dispersion observed in
CsH$_2$PO$_4$.  It should be mentioned that similar studies of thermodynamic and dynamic characteristics of  KH$_2$PO$_4$  can be
carried out using the technique developed in~\cite{238pok}.
 Such studies would make it possible  to explore the
effect of suppression of the soft mode in the KH$_2$PO$_4$ type
ferroelectrics more consistently than in~\cite{48pok} and thereby to
explain the Debye character of the dielectric permittivity
dispersion in these crystals.

It should be noted that the ferroelectric compounds of the
 KH$_2$PO$_4$ family are piezoelectric. Piezoelectric coupling is observed in external electric fields and mechanical stresses of certain symmetries.
Ferroelectric phase transition in the  KH$_2$PO$_4$ type crystals is accompanied by the appearance of spontaneous strains, which changes their
tetragonal symmetry. So far, the calculations of dielectric characteristics of these crystals within the proton ordering model
\cite{145x,58x,342x,343x,344x,163x,165x,uni2009} were restricted to a static limit and high-frequency relaxation. The attempts to explore the piezoelectric
resonance phenomenon within a model that does not take into account the piezoelectric coupling were vain.  The conventional proton ordering model does not permit one to describe the effects associated with the differences of the free and clamped crystal regimes in the static limit or
the phenomenon of  crystal clamping by a high-frequency field. This leads, in particular, to some quantitative deviations from experiment for the
temperature behavior of polarization relaxation time and dynamic dielectric permittivity of the  KH$_{2}$PO$_{4}$ type ferroelectrics in the phase transition region.

The studies of the piezoelectric coupling effect on the phase transition and on physical characteristics of the  KH$_2$PO$_4$ type ferroelectrics were initiated in\cite{0311U1}, where the Slater theory~\cite{0311U2} was modified by taking into account
the splitting of the lowest ferroelectric level due to the strain $\varepsilon_6$.

The most fundamental results for the KH$_{2}$PO$_{4}$ family ferroelectrics were obtained in~\cite{0311U4,0311U5,0311U6,Lis2003,JPS1701,0311U8,0311U7,lis2007,Stasyuk2008}. For the deformed crystals of the KH$_{2}$PO$_{4}$ type,
 the Hamiltonian of the proton ordering model was modified for the first time by including  $\varepsilon _{6}$ into in the shear strain \cite{0311U4,0311U5}, taking into account the deformational mean field and the splitting of the lateral proton configurations. Later on \cite{0311U6,Lis2003}, all possible splittings of proton configurational energies by the strain $\varepsilon
_{6}$ were included into the model. In~\cite{0311U6}, using this model, the phase transition, thermodynamic and longitudinal dielectric, piezoelectric, and elastic characteristics of
K(H$_{0.12}$D$_{0.88})_{2}$PO$_{4 }$, as well as the effect of the  $\sigma
_{6 }$ on these quantities were explored. The same characteristics for other K(H$_{1-x}$D$_{x})_{2}$PO$_{4}$ type ferroelectrics were later on calculated in \cite{JPS1701}.

The thermodynamic and longitudinal dielectric, piezoelectric, and elastic characteristics
of the KH$_{2}$PO$_{4}$ type were also calculated in~\cite{Lis2003,0311U8} within a model that takes into account the tunneling and piezoelectric coupling. It should be mentioned, however, that taking into account the tunneling within the cluster approximation yields a non-physical behavior of the calculated quantities at low temperatures~\cite{471x}. In~\cite{0311U7,lis2007,Stasyuk2008},  the effect of the electric field $E_{3}$ on the phase transition and on the physical characteristics of
K(H$_{0.12}$D$_{0.88})_{2}$PO$_{4}$ and KH$_{2}$PO$_{4}$ was explored, and a good agreement with experiment was obtained.

In~\cite{0311U6,Lis2003,JPS1701,0311U8,0311U7}, where there was used a model with tunneling, the dynamic characteristics of the  KH$_{2}$PO$_{4}$ family ferroelectrics were not considered.
In~\cite{cmp555}, using a modified proton ordering model  proposed in~\cite{0311U6}, the dynamic dielectric permittivity of a free KH$_{2}$PO$_{4}$ type crystals was calculated taking into account the dynamics of  $\varepsilon_6$ strain. The experimentally observed effects of crystal clamping by a high-frequency electric field and  piezoelectric resonance in KH$_2$PO$_4$ and KD$_2$PO$_4$ crystals were theoretically described for the first time. Peculiarities of the ultrasound attenuation coefficients near the phase transition temperature in these crystals were also described.
In~\cite{533x}, we presented a detailed review of the obtained results for
longitudinal and transverse static dielectric permittivities, for piezoelectric coefficients, and for elastic constants of
several ferroelectric crystals of the  KH$_2$PO$_4$ family. Moreover, the typical behavior of longitudinal and transverse
characteristics of mechanically free KH$_2$PO$_4$, Rb$_2$PO$_4$, KH$_2$AsO$_4$ crystals was shown and  the results for
temperature and frequency dependencies of longitudinal and transverse dielectric permittivities of KH$_2$PO$_4$ were presented, along with the
corresponding experimental data.

In the present paper, using the  model proposed in~\cite{0311U6} we calculate the longitudinal dynamic dielectric permittivity of
clamped ferroelectrics of the  KH$_2$PO$_4$ type and explore its behavior in wide temperature and frequency ranges.
Using the obtained results, we perform a detailed analysis of the available experimental data for these crystals.

\section{Systems of equations for the time-dependent deuteron distribution functions}

\begin{figure}[!ht]
\begin{center}
 \includegraphics[scale=0.6]{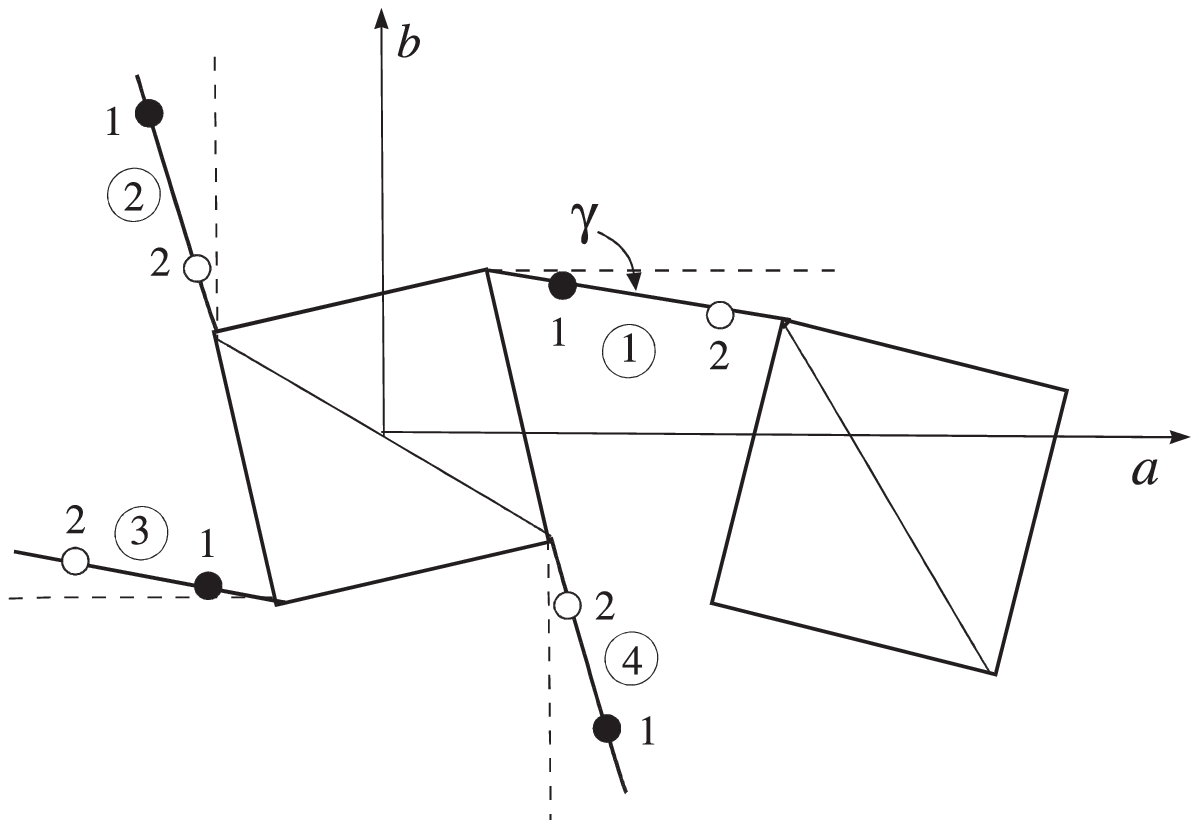}
\end{center}
\caption{A primitive cell of the  KD$_2$PO$_4$ type crystal. One of the numerous possible ferroelectric deuteron configurations is shown.} \label{KDP2}
\end{figure}

We shall consider a system of deuterons moving on the  O--D{\ldots}O bonds
in deuterated KD$_2$PO$_4$ type ferroelectrics. A
primitive cell of the Bravais lattice of these crystals consists of
two neighboring tetrahedra PO$_4$ along with four hydrogen bonds
attached to one of them (the ``A'' type tetrahedron). The hydrogen
bonds attached to the other tetrahedron (``B'' type) belong to the four
 structural elements surrounding this tetrahedron (figure~\ref{KDP2}).

The dynamic characteristics of these compounds will be calculated within the four-particle cluster approximation that proved to be successful in describing their thermodynamic properties~\cite{0311U6,0311U8,JPS1701,Lis2003,uni2009}.

The Hamiltonian of the deuteron subsystem, taking into account short-range and
long-range interactions in the presence of an external electric field $E_3$ along the crystallographic $c$ axis and mechanical stress $\sigma_6 = \sigma_{xy}$, which independently contribute to polarization $P_3$ and strain
 $\varepsilon_6$, consists of the ``seed'' and pseudospin parts~\cite{0311U6,JPS1701}:
 \be
 \hat H = NH^{(0)} + \hat H_\mathrm{s}\, ,
 \ee
where $N$ is the total number of primitive cells. The ``seed''
energy of a primitive cell corresponds to the sublattice of heavy ions and does not explicitly depend
on the deuteron subsystem configuration. It is expressed in terms of the strain $\varepsilon_6$ and electric field
 $E_3$ and includes the elastic, piezoelectric, and dielectric contributions
 \be
 H^{(0)} = {v} \left( \frac12 c_{66}^{E0}\varepsilon_6^2 -
 e_{36}^0E_3\varepsilon_6 - \frac12 \chi_{33}^{\varepsilon 0}E_3^2
 \right),
 \ee
where $v$ is the
primitive cell volume; $c_{66}^{E0}$, $e_{36}^0$,
 $\chi_{33}^{\varepsilon 0}$ are the ``seed'' elastic constant,
piezoelectric coefficient, and dielectric susceptibility, respectively. They determine the temperature behavior of
the corresponding observable quantities at temperatures far from the phase transition $T_\mathrm{c}$.

The pseudospin part of the Hamiltonian reads
 \bea
 \hat H_\mathrm{s} = \frac12 \sum\limits_{ {qf}\atop{q'f'}}J_{ff'}(qq')
 \frac{\sigma_{qf}}{2}\frac{\sigma_{q'f'}}{2} + \hat H_\mathrm{sh.s}(6)+
  \sum\limits_{qf} 2\psi_6\varepsilon_6 \frac{\sigma_{qf}}{2}
 - \sum\limits_{qf}\mu_{f3}E_3 \frac{\sigma_{qf}}{2}\,.
 \eea
The first term describes effective long-range interactions between
deuterons; $\sigma_{qf}$ is the $z$-th component of the pseudospin
operator that describes the state of a deuteron in
the $q$-th cell on the $f$-th bond. ($f=1,2,3,4$). Two eigenvalues of the operator
$\sigma_{qf} = \pm 1$ correspond to two possible positions of the
deuteron on the bond denoted by ``1'', ``2'' in figure~\ref{KDP2}.
In (2.3) $\hat H_\mathrm{sh.s}(6)$ is a linear over the strain
 $\varepsilon_6$ Hamiltonian of the short-range interactions between deuterons~\cite{JPS1701}:
 \bea
 \hat H_\mathrm{sh.s}(6)& =& \sum\limits_q    \Bigg[   \left(
 \frac{\delta_{s6}}{8}\varepsilon_6 -
 \frac{\delta_{16}}{4}\varepsilon_6\right) \left(\sigma_{q1} + \sigma_{q2} + \sigma_{q3} +
 \sigma_{q4}\right)\non
& &{} {-}   \left( \frac{\delta_{s6}}{8}\varepsilon_6 + \frac{\delta_{16}}{4}\varepsilon_6\right)
 \left(\sigma_{q1} \sigma_{q2} \sigma_{q3} + \sigma_{q1} \sigma_{q2} \sigma_{q4}
 + \sigma_{q1} \sigma_{q3} \sigma_{q4} + \sigma_{q2} \sigma_{q3}
 \sigma_{q4}\right)  \non
 & &{}+  \frac14 \left(V_\mathrm{s} + \delta_{a6}\varepsilon_6\right)\left(\sigma_{q1} \sigma_{q2} + \sigma_{q3} \sigma_{q4}
 \right)+ \frac14 \left(V_\mathrm{s} - \delta_{a6}\varepsilon_6\right)\left(\sigma_{q2} \sigma_{q3} + \sigma_{q4}
 \sigma_{q1}\right)  \nonumber\\
 &&{} +  \frac14 U_\mathrm{s} \left(\sigma_{q1} \sigma_{q3} + \sigma_{q2}
 \sigma_{q4}\right) + \frac{1}{16} \Phi_\mathrm{s} \sigma_{q1} \sigma_{q2} \sigma_{q3}
 \sigma_{q4}\Bigg]\,.
 \eea
Here
 \[
 V_\mathrm{s} = - \frac12 w_1, \qquad U_\mathrm{s} = \frac12 w_1 - \varepsilon, \qquad \Phi_\mathrm{s} =
 4\varepsilon- 8w + 2w_1
 \]
 and
 \[
 \varepsilon = \varepsilon_a - \varepsilon_s, \qquad  w = \varepsilon_1 -
 \varepsilon_s, \qquad w_1 = \varepsilon_0 - \varepsilon_s\, ,
 \]
where $\varepsilon_s$, $\varepsilon_a$, $\varepsilon_1$,
$\varepsilon_0$ are the energies of deutron configurations near the
PO$_4$ group.

The third term in  (2.3) is a linear over the shear strain
$\varepsilon_6$ mean field Hamiltonian induced by the piezoelectric coupling; $\psi_6$ is the parameter
of the deformational mean field.

The last term in  (2.3) effectively describes the interactions of deuterons
with an external electric field $E_3$. Here $\mu_{f3}$ is the effective dipole moment related to the  $f$-th
hydrogen bonds, where
 \[
 \mu_{13} = \mu_{23} = \mu_{33} = \mu_{43} = \mu_{3}\,,
 \]
and $\mu_{3}$ is the dipole moment of up/down deuteron configurations.

Taking into account the peculiarities of the crystalline structure of the
MD$_2$XO$_4$ the type ferroelectrics, their dynamic characteristics can be calculated within the four-particle cluster approximation that proved to be effective in describing the thermodynamic characteristics of these crystals~\cite{uni2009,0311U6,Lis2003,JPS1701,0311U8}.
Long-range interactions are taken into account in the mean field approximation. Within the cluster approach, the thermodynamic
 potential of  MD$_2$XO$_4$ ferroelectrics calculated per one primitive cell reads
\bea  && g_\mathrm{s}(6) =  H^{(0)}
+ 2\nu_\mathrm{c}\left(\eta^{(1)z}\right)^2 + \frac{1}{2\beta}\sum\limits_{f=1}^4\ln
 Z_{f\mathrm{s}}^{(1)} - \frac1\beta\ln Z_{6\mathrm{s}}^{(4)} - {{v}}\sigma_6\varepsilon_6\,,
\eea
where $4\nu_\mathrm{c} = J_{11}(0) + 2J_{12}(0) + J_{13}(0)$, the eigenvalues of Fourier-transform
of the long-range interaction matrix
 $J_{ff'} = \sum\limits_{ {\bf R}_q - {\bf R}_{q'}} J_{ff'}(qq')$;
 \[
  \eta^{(1)z} = \langle \sigma_{q1} \rangle = \langle \sigma_{q2} \rangle
 = \langle \sigma_{q3} \rangle = \langle \sigma_{q4} \rangle
 \]
 is the parameter of deuteron ordering; $Z_{f\mathrm{s}}^{(1)}=\mathrm{Sp}\, \re^{-\beta\hat H_{qf\mathrm{s}}^{(1)}}$,
$Z_{6\mathrm{s}}^{(4)}=\mathrm{Sp}\, \re^{-\beta\hat H_{q\mathrm{s}}^{(4)} }$, $\beta=\frac{1}{k_\mathrm{B}T}$ are the single-particle and four-particle
partition functions.
The single-particle $\hat H_{qf\mathrm{s}}^{(1)}$ and four-particle $\hat
H_{q\mathrm{s}}^{(4)}$ deuteron Hamiltonians are presented  by
 \be
 \hat H_{qf\mathrm{s}}^{(1)} = - \frac{\bar z_{6}}{\beta}\frac{\sigma_{qf}}{2}\,,
 \ee
 \bea
 && \hat H_{q\mathrm{s}}^{(4)} = - \sum\limits_{f=1}^4 \frac{z_6}{\beta}
 \frac{\sigma_{qf}}{2}+ \hat H_\mathrm{sh.s}(6),
  \eea
where
 \[
 z_6 = \beta \left(- \Delta_\mathrm{s}^\mathrm{c} + 2\nu_\mathrm{c} \eta^{(1)z} - 2\psi_6\varepsilon_6 +\mu_3E_3\right), \qquad \bar z_{6} = \beta \left(- 2\Delta_\mathrm{s}^\mathrm{c}  + 2\nu_\mathrm{c} \eta^{(1)z} - 2\psi_6\varepsilon_6 + \mu_3E_3 \right).
 \]
The effective field $\Delta_\mathrm{s}^\mathrm{c}$ exerted by the neighboring hydrogen bonds from
outside the cluster,
is determined from the self-consistency condition: the
mean values $\langle \sigma_{qf} \rangle$ calculated within the
four-particle and one-particle cluster approximations should
coincide.

The dynamic characteristics of the MD$_{2}$XO$_{4}$
crystals will be explored using the proposed dynamic
model  based on a stochastic Glauber
model~\cite{61x}. Using the method developed in~\cite{163x,165x,uni2009,cmp555}, the
system of equations for the time-dependent deuteron distribution
functions is obtained in the form
  \be
  - \alpha \frac{\rd}{\rd t} \left\langle \prod_f \sigma_{qf} \right\rangle =
  \sum\limits_{f'} \left\{ \left\langle \prod_f \sigma_{qf} \left[ 1 -
  \sigma_{qf'} \tanh \frac12 \beta {\mbox{\boldmath$\varepsilon$}}_{qf'}^z(t)
  \right] \right\rangle \right\},
  \ee
 where ${\mbox{\boldmath$\varepsilon$}}_{qf'}^z(t)$  is the local field acting on the  $f'$-th
deuteron in the  $q$-th cell, which can be obtained from the Hamiltonian  (2.3). Expanding $ \tanh \frac12 \beta {\mbox{\boldmath$\varepsilon$}}_{qf'}^z(t)$ over the pseudospin operators $\sigma_{qf}$, occurring in Hamiltonian (2.3), taking into account the fact that $\sigma_{qf}=\pm 1$ and the symmetry of the deuteron distribution functions in the   MD$_2$XO$_4$ ferroelectrics in the presence of the electric field $ E_3$
 \bea
  \eta^{(1)z} &=& \langle \sigma_{q1} \rangle =  \langle \sigma_{q2}
 \rangle =  \langle \sigma_{q3} \rangle =  \langle \sigma_{q4}
 \rangle, \non
 \eta^{(3)z}& =& \langle \sigma_{q1}  \sigma_{q2} \sigma_{q3} \rangle =  \langle \sigma_{q1}
 \sigma_{q3} \sigma_{q4}  \rangle =  \langle \sigma_{q1}  \sigma_{q2} \sigma_{q4}
 \rangle = \langle \sigma_{q2}  \sigma_{q3} \sigma_{q4} \rangle,\nonumber \\
 \eta_1^{(2)z} &=& \langle \sigma_{q2} \sigma_{q3} \rangle \!=\! \langle \sigma_{q1} \sigma_{q4}
 \rangle,
 \qquad \eta_2^{(2)z} \!=\! \langle \sigma_{q1} \sigma_{q2} \rangle \!=\! \langle \sigma_{q3} \sigma_{q4}
 \rangle,
 \qquad \eta_3^{(2)z} \!=\! \langle \sigma_{q1} \sigma_{q3} \rangle \!=\! \langle \sigma_{q2} \sigma_{q4}
 \rangle, \quad
 \eea
from (2.7), one can obtain a closed system of equations for the time-dependent single-particle, three-particle, and pair distribution functions of
deuterons in  MD$_2$XO$_4$ within the four-particle cluster approximation and for a single-particle distribution function within
the single-particle approximation~\cite{cmp555}:
\be
 \alpha \frac{\rd}{\rd t} \left(\!\! \begin{array}{ccccc}
                        \eta^{(1)z}\\
                        \eta^{(3)z}\\
                        \eta_1^{(2)z}\\
                        \eta_2^{(2)z}\\
                        \eta_3^{(2)z}
                       \end{array}
                        \!\!\right) \!=\!  \left(\!\! \begin{array}{ccccc}
                                \bar c_{11} & \bar c_{12} & \bar c_{13} & \bar c_{14} & \bar c_{15}  \\
                                \bar c_{21} & \bar c_{22} & \bar c_{23} & \bar c_{24} & \bar c_{25}  \\
                                \bar c_{31} & \bar c_{32} & \bar c_{33} & \bar c_{34} & \bar c_{35}  \\
                                \bar c_{41} & \bar c_{42} & \bar c_{43} & \bar c_{44} & \bar c_{45}  \\
                                \bar c_{51} & \bar c_{52} & \bar c_{53} & \bar c_{54} & \bar c_{55}
                                \end{array}
                                \!\!\right)
                        \left(\!\! \begin{array}{ccccc}
                        \eta^{(1)z}\\
                        \eta^{(3)z}\\
                        \eta_1^{(2)z}\\
                        \eta_2^{(2)z}\\
                        \eta_3^{(2)z}
                        \end{array}
                        \!\!\right) \!+\!
                                \left(\!\! \begin{array}{ccccc}
                                \bar c_1 \\
                                \bar c_2 \\
                                \bar c_3 \\
                                \bar c_4 \\
                                \bar c_5
                                \end{array}
                                \!\!\right),
 \label{deta14}\ee
where the following notations are used
 \bea
 && \hspace{-5ex} \bar c_{11} = - (1 - P_6^z - Q_{61}^z - Q_{62}^z), \, \quad\qquad \bar c_{12}
 =R_6^z,  \hspace{2.5mm} \qquad \bar c_{13} = M_{61}^z, \qquad \bar c_{14} = M_{62}^z, \qquad \bar c_{15} =
 N_{6}^z, \qquad \bar c_1 = L_6^z, \non
 && \hspace{-5ex} \bar c_{21} = (2P_6^z + 2Q_{61}^z + 2Q_{62}^z +
 3R_6),\quad \bar c_{22} = - (3 - P_6^z - Q_{61}^z - Q_{62}^z), \quad \bar c_{23} = (N_6^z + M_{62}^z + L_6^z), \non
 && \hspace{-5ex}  \bar c_{24} = (N_6^z + M_{61}^z +
 L_6^z), \hspace{17mm} \bar c_{25} = (M_{61}^z + M_{62}^z + L_6^z),  \hspace{10.5mm} \bar c_2 = (N_6^z + M_{61}^z + M_{62}^z),\non
 && \hspace{-5ex} \bar c_{31} = 2(N_6^z + M_{62}^z +
 L_6^z), \qquad \bar c_{32} = 2M_{61}^z, \qquad \bar c_{33} = - 2(1 - R_6^z), \qquad \bar c_{34} = 2P_6^z, \qquad \bar c_{35}
 = 2Q_{61}^z, \qquad \bar c_{3} = 2Q_{62}^z, \non
 && \hspace{-5ex} \bar c_{41} = 2(N_6^z + M_{61}^z + L_6^z), \qquad \bar c_{42} =
 2M_{62}^z, \qquad \bar c_{43} = 2P_6^z, \qquad \bar c_{44} = -2 (1 -
 R_6^z), \qquad \bar c_{45} = 2Q_{62}^z,\qquad \bar c_{4} = 2Q_{61}^z, \non
 && \hspace{-5ex}  \bar c_{51} = 2(M_{61}^z + M_{62}^z + L_6^z), \hspace{4.5mm} \bar c_{52} = 2N_6^z, \hspace{7.5mm} \bar c_{53} = 2Q_{61}^z, \hspace{4.5mm} \bar c_{54} = 2Q_{62}^z, \qquad \bar c_{55}
 = - 2(1 - R_6^z), \qquad \bar c_{5} = 2P_6^z, \non
 \eea
 \be
 \alpha \frac{\rd}{\rd t} \eta^{(1)z} = - \eta^{(1)z} + \tanh
 \frac12 \bar z_6\,. \label{deta1}
 \ee

 \section{Relaxational dynamics of mechanically clamped MD$_2$XO$_4$ crystals}

Now, using the obtained systems of equations,  let us calculate the dynamic characteristics of the MD$_2$XO$_4$ crystals.
 Let us consider the case of small deviations of the considered
system from equilibrium. We can separate the static and dynamic parts in the obtained system of
equations. To do so, we present the
distribution functions and the effective fields as sums of the
equilibrium functions and their fluctuations
 \bea
 && \eta^{(1)z} =  \eta^{(1)} + \eta_t^{(1)z}, \qquad
 \eta^{(3)z} =  \eta^{(3)} + \eta_t^{(3)z},
 \qquad \eta_{i}^{(2)z} =  \eta_i^{(2)} + \eta_{it}^{(2)z}
 \qquad (i=1,2,3), \nonumber \\
 && z_6 =  \tilde{z_6} + z_{6t}\, ,\qquad \tilde{z_6}  = - \beta \tilde \Delta^{c}_\mathrm{s} + 2\beta
 \nu_c  \eta^{(1)} - 2 \beta \psi_6\varepsilon_6\, , \qquad z_{6t}= - \beta \Delta^{c}_{st} + 2\beta\nu_c \eta_t^{(1)z} + \beta\mu_3E_{3t}\, .\qquad
 \eea
Owing to a piezoelectric coupling, time-dependent electric fields should induce time-dependent
strains. However, in the present paper we shall consider the fields with the frequencies of the order of  $10^9 \sim
 10^{12}$~Hz, which is far above the frequency of piezoelectric resonance. When the frequency is that high, the strains are not capable  of following the external fields, which means that the crystal is effectively clamped. Therefore, in the expansions (3.1) we assume the strain $\varepsilon_6$ to be time-independent.

We expand the expressions for the coefficients  $P_6^z, \dots, L_6^z$  in series in
 $z_{6t}/2$  up to the linear
terms. Taking into account these expansions and (3.1), we obtain a system of equations that  describes the behavior of fluctuational parts of distribution functions~\cite{cmp555,icmp0608U}:
 \be
 \frac{\rd}{\rd t} \left( \begin{array}{ccccc}
                        \eta_t^{(1)z} \\
                        \eta_t^{(3)z} \\
                        \eta_{1t}^{(2)z}\\
                        \eta_{2t}^{(2)z}\\
                        \eta_{3t}^{(2)z}
                        \end{array}
                        \right) = \left( \begin{array}{ccccc}
                                   c_{11} &  c_{12} &  c_{13} &  c_{14} & c_{15} \\
                                   c_{21} &  c_{22} &  c_{23} &  c_{24} & c_{25} \\
                                   c_{31} &  c_{32} &  c_{33} &  c_{34} & c_{35}  \\
                                   c_{41} &  c_{42} &  c_{43} &  c_{44} & c_{45}  \\
                                   c_{51} &  c_{52} &  c_{53} &  c_{54} & c_{55}
                                \end{array}
                                \right) \left( \begin{array}{ccccc}
                        \eta_t^{(1)z} \\
                        \eta_t^{(3)z} \\
                        \eta_{1t}^{(2)z}\\
                        \eta_{2t}^{(2)z}\\
                        \eta_{3t}^{(2)z}
                        \end{array}
                        \right) - \frac{\mu_3 E_{3t}}{2kT}
                                \left( \begin{array}{ccccc}
                                c_1 \\
                                c_2 \\
                                c_3 \\
                                c_4 \\
                                c_5
                                \end{array}
                                \right),
 \ee
 where the coefficients of the system read
\begin{align*}
 & c_{11} = \frac{1}{\alpha} \Bigl( \bar c_{11}^{(0)} + \beta
 \nu_c Y_s^{(1)} - k_s^{(1)} \Psi_s^z\Bigr), &
 & c_{12} = \frac{1}{\alpha} \Bigl( \bar c_{12}^{(0)} - k_s^{(1)}\bar
 c_{12}^{(0)}\Bigr), &
 & c_{13} = \frac{1}{\alpha} \Bigl( \bar c_{13}^{(0)} - k_s^{(1)}\bar
 c_{13}^{(0)}\Bigr),  \non
 & c_{14} = \frac{1}{\alpha} \Bigl( \bar c_{14}^{(0)} - k_s^{(1)}\bar
 c_{14}^{(0)}\Bigr), &
 & c_{15} = \frac{1}{\alpha} \Bigl( \bar c_{15}^{(0)} - k_s^{(1)}\bar
 c_{15}^{(0)}\Bigr), &
 & c_1 = \frac{1}{\alpha}k_s^{(1)} r_s\,, \non
 & c_{21} =\frac{1}{\alpha} \Bigl( \bar c_{21}^{(0)} + \beta
 \nu_c Y_s^{(3)} - k_s^{(3)} \Psi_s^z\Bigr),&
 & c_{22} = \frac{1}{\alpha} \Bigl( \bar c_{22}^{(0)} - k_s^{(3)}\bar
 c_{12}^{(0)}\Bigr), &
 & c_{23} = \frac{1}{\alpha} \Bigl( \bar c_{23}^{(0)} - k_s^{(3)}\bar
 c_{13}^{(0)}\Bigr), \non
 & c_{24} = \frac{1}{\alpha} \Bigl( \bar c_{24}^{(0)} - k_s^{(3)}\bar
 c_{14}^{(0)}\Bigr), &
 & c_{25} = \frac{1}{\alpha} \Bigl( \bar c_{25}^{(0)} - k_s^{(3)}\bar
 c_{15}^{(0)}\Bigr), &
 & c_2 = \frac{1}{\alpha}k_s^{(3)} r_s\,, \non
\end{align*}
\begin{align*}
 & c_{31} =\frac{1}{\alpha} \Bigl( \bar c_{31}^{(0)} + \beta
 \nu_c Y_{s1}^{(2)} - k_{s1}^{(2)} \Psi_s^z\Bigr), &
 & c_{32} = \frac{1}{\alpha} \Bigl( \bar c_{32}^{(0)} - k_{s1}^{(2)}\bar
 c_{12}^{(0)}\Bigr), &
 & c_{33} = \frac{1}{\alpha} \Bigl( \bar c_{33}^{(0)} - k_{s1}^{(2)}\bar
 c_{13}^{(0)}\Bigr),
 \non
 &c_{34} = \frac{1}{\alpha} \Bigl( \bar c_{34}^{(0)} - k_{s1}^{(2)}\bar
 c_{14}^{(0)}\Bigr),&
 & c_{35} = \frac{1}{\alpha} \Bigl( \bar c_{35}^{(0)} - k_{s1}^{(2)}\bar
 c_{15}^{(0)}\Bigr), &
 & c_3 = \frac{1}{\alpha}k_{s1}^{(2)} r_s\,, \non
 & c_{41} =\frac{1}{\alpha} \Bigl( \bar c_{41}^{(0)} + \beta
 \nu_c Y_{s2}^{(2)} - k_{s2}^{(2)} \Psi_s^z\Bigr),&
 & c_{42} = \frac{1}{\alpha} \Bigl( \bar c_{42}^{(0)} - k_{s2}^{(2)}\bar
 c_{12}^{(0)}\Bigr), &
 & c_{43} = \frac{1}{\alpha} \Bigl( \bar c_{43}^{(0)} - k_{s2}^{(2)}\bar
 c_{13}^{(0)}\Bigr), \non
 & c_{44} = \frac{1}{\alpha} \Bigl( \bar c_{44}^{(0)} - k_{s2}^{(2)}\bar
 c_{14}^{(0)}\Bigr), &
 & c_{45} = \frac{1}{\alpha} \Bigl( \bar c_{45}^{(0)} - k_{s2}^{(2)}\bar
 c_{15}^{(0)}\Bigr), &
 & c_4 = \frac{1}{\alpha}k_{s2}^{(2)} r_6\,,
 \non
 & c_{51} =\frac{1}{\alpha} \Bigl( \bar c_{51}^{(0)} + \beta
 \nu_c Y_{s3}^{(2)} - k_{s3}^{(2)} \Psi_s^z\Bigr), &
 &
 c_{52} = \frac{1}{\alpha} \Bigl( \bar c_{52}^{(0)} - k_{s3}^{(2)}\bar
 c_{12}^{(0)}\Bigr), &
 & c_{53} = \frac{1}{\alpha} \Bigl( \bar c_{53}^{(0)} - k_{s3}^{(2)}\bar
 c_{13}^{(0)}\Bigr),  \non
 & c_{54} = \frac{1}{\alpha} \Bigl( \bar c_{54}^{(0)} - k_{s3}^{(2)}\bar
 c_{14}^{(0)}\Bigr), &
 & c_{55} = \frac{1}{\alpha} \Bigl( \bar c_{55}^{(0)} - k_{s3}^{(2)}\bar
 c_{15}^{(0)}\Bigr), &
 & c_5 = \frac{1}{\alpha}k_{s3}^{(2)} r_s\, .
\end{align*}
 The expressions for the quantities entering the coefficients of the system (3.2), are given in~\cite{icmp0608U}; if the
 piezoelectric coupling is neglected, they coincide with the corresponding expressions of~\cite{uni2009}.

The system of equations (3.2) is reduced to a non-uniform differential equation with constant coefficients
  for a single-particle distribution function
 \bea
   \stackrel{.....}{\eta}_t^{(1)z} +
 p_4\stackrel{....}{\eta}_t^{(1)z}+
 p_3\stackrel{...}{\eta}_t^{(1)z}+
 p_2\stackrel{..}{\eta}_t^{(1)z}
 + p_1\stackrel{.}{\eta}_t^{(1)z} + p_0\eta_t^{(1)z} =
 \frac{\mu_3 E_{3t}}{2}\beta p,
 \eea
where  $p = - \left[(\ri\omega)^4 p^{(4)} + (\ri\omega)^3 p^{(3)} + (\ri\omega)^2 p^{(2)} + (\ri\omega) p^{(1)} +
 p^{(0)}\right]$.
 Expressions for coefficients  $p_4,\ldots,p^{(0)}$ are presented in~\cite{icmp0608U}.

Finally, time-dependent single-particle distribution function is obtained in the following form
 \be
 \eta_t^{(1)z} = \sum\limits_{i=1}^5 C_i^z \exp \left( -
 \frac{t}{\tau_i^z} \right) + \frac{\mu_3 E_{3t}}{2}\beta
 \frac{\sum\limits_{k=0}^4 (\ri\omega)^kp^{(k)}}{(\ri\omega)^5 + \sum\limits_{k=0}^4 (\ri\omega)^k
 p_k}\,. \label{etat}
 \ee
 Here $C_i^z$ are constant coefficients;  $\tau_i^z$  are relaxation times represented by
 \[
 \tau_i^z = (- q_i^z)^{-1},
 \]
 where $q_i^z$  are roots of the characteristics equation
 \be
 (q^z)^5 + p_4(q^z)^4 + p_3(q^z)^3 + p_2(q^z)^2 + p_1(q^z)
 + p_0 = 0.
 \ee

The dynamic dielectric susceptibility of a clamped crystal is defined as
\bea
\chi_{33}^{\varepsilon}(\omega, T) = \lim \limits_{E_{3t} \to 0}
 2 \frac{\mu_3}{v} \frac{\rd \eta_t^{(1)z}}{\rd E_{3t}}= \frac{\mu_3^2}{v}
 \beta \frac{\sum\limits_{k=0}^4 (\ri\omega)^kp^{(k)} }
 {(\ri\omega)^5 + \sum\limits_{k=0}^4 (\ri\omega)^k p_k}
  =  \frac{\mu_3^2}{v} \beta
 \frac{\prod\limits_{i=1}^5 \tau_i^z \Bigl[ \sum\limits_{k=0}^4 (\ri\omega)^kp^{(k)}\Bigr] }
 {\prod\limits_{i=1}^5 (1+ \ri\omega\tau_i^z)} = \sum\limits_{i=1}^5
 \frac{\chi_{3i}}{1+ \ri\omega\tau_i^z }\,.
\eea
The coefficients $\chi_{3i}$ are found from the following system of equations
\be
 \left( \begin{array}{ccccc}
         n_{11} & n_{12} & n_{13} & n_{14} & n_{15}  \\
         n_{21} & n_{22} & n_{23} & n_{24} & n_{25} \\
         n_{31} & n_{32} & n_{33} & n_{34} & n_{35} \\
         n_{41} & n_{42} & n_{43} & n_{44} & n_{45} \\
         n_{51} & n_{52} & n_{53} & n_{54} & n_{55}
                        \end{array}
                        \right)
                             \left( \begin{array}{c}
                                    \chi_{31} \\
                                    \chi_{32} \\
                                    \chi_{33} \\
                                    \chi_{34} \\
                                    \chi_{35}
                                    \end{array}
                                    \right) = \left( \begin{array}{c}
                                                n_1 \\
                                                n_2 \\
                                                n_3 \\
                                                n_4 \\
                                                n_5
                                                \end{array}
                                                \right).
 \ee
Here,  the following notations are used
%
\begin{align*}
 & n_{11} = \tau_2^z\tau_3^z\tau_4^z\tau_5^z; ~~~~~~ n_{12} =
 \tau_1^z\tau_3^z\tau_4^z\tau_5^z; ~~~~ n_{13} =
 \tau_1^z\tau_2^z\tau_4^z\tau_5^z; &
 & n_{14} = \tau_1^z\tau_2^z\tau_3^z\tau_5^z; ~~~~ n_{15} =
 \tau_1^z\tau_2^z\tau_3^z\tau_4^z;\non
 & n_{21} = \tau_2^z\tau_3^z\tau_4^z + \tau_2^z\tau_3^z\tau_5^z +
 \tau_2^z\tau_4^z\tau_5^z + \tau_3^z\tau_4^z\tau_5^z; &
 & n_{22} = \tau_1^z\tau_3^z\tau_5^z + \tau_1^z\tau_4^z\tau_5^z +
 \tau_1^z\tau_3^z\tau_4^z + \tau_3^z\tau_4^z\tau_5^z; \non
 & n_{23} = \tau_1^z\tau_2^z\tau_4^z + \tau_1^z\tau_2^z\tau_5^z +
 \tau_1^z\tau_4^z\tau_5^z + \tau_2^z\tau_4^z\tau_5^z;&
 & n_{24} = \tau_1^z\tau_2^z\tau_3^z + \tau_1^z\tau_2^z\tau_5^z +
 \tau_1^z\tau_3^z\tau_5^z + \tau_2^z\tau_3^z\tau_5^z; \non
 & n_{25} = \tau_1^z\tau_2^z\tau_3^z + \tau_1^z\tau_2^z\tau_4^z +
 \tau_1^z\tau_3^z\tau_4^z + \tau_2^z\tau_3^z\tau_4^z;&
 & n_{31} = \tau_2^z\tau_3^z + \tau_2^z\tau_4^z +
 \tau_3^z\tau_4^z + \tau_2^z\tau_5^z + \tau_3^z\tau_5^z +
 \tau_4^z\tau_5^z; \non
 & n_{32} = \tau_1^z\tau_3^z + \tau_1^z\tau_4^z +
 \tau_1^z\tau_5^z + \tau_3^z\tau_4^z + \tau_3^z\tau_5^z +
 \tau_4^z\tau_5^z; &
 & n_{33} = \tau_1^z\tau_2^z + \tau_1^z\tau_4^z +
 \tau_1^z\tau_5^z + \tau_2^z\tau_4^z + \tau_2^z\tau_5^z +
 \tau_4^z\tau_5^z; \non
 & n_{34} = \tau_1^z\tau_2^z + \tau_1^z\tau_3^z +
 \tau_1^z\tau_5^z + \tau_2^z\tau_3^z + \tau_2^z\tau_5^z +
 \tau_3^z\tau_5^z; &
 & n_{35} = \tau_1^z\tau_2^z + \tau_1^z\tau_3^z +
 \tau_1^z\tau_4^z + \tau_2^z\tau_3^z + \tau_2^z\tau_4^z +
 \tau_3^z\tau_4^z; \non
 & n_{41} = \tau_2^z + \tau_3^z + \tau_4^z + \tau_5^z, \qquad \
  n_{42} = \tau_1^z + \tau_3^z + \tau_4^z + \tau_5^z; &
 & n_{43} = \tau_1^z + \tau_2^z + \tau_4^z + \tau_5^z, \non
 & n_{44} = \tau_1^z + \tau_2^z + \tau_3^z + \tau_5^z, \qquad \
  n_{45} = \tau_1^z + \tau_2^z + \tau_3^z + \tau_4^z; &
 & n_{51} = n_{52} = n_{53} = n_{54} = n_{55} = 1;
 \end{align*}
 \begin{eqnarray}
&  n_1 = \frac{\mu_3^2}{v} \beta \prod\limits_{i=1}^5
 \tau_i^z p^{(4)}, \qquad n_2 = \frac{\mu_3^2}{v} \beta \prod\limits_{i=1}^5
 \tau_i^z p^{(3)}, \qquad   n_3 = \frac{\mu_3^2}{v} \beta \prod\limits_{i=1}^5
 \tau_i^z p^{(2)}, \nonumber\\
 & n_4 = \frac{\mu_3^2}{v}\beta  \prod\limits_{i=1}^5
 \tau_i^z p^{(1)}, \qquad n_5 = \frac{\mu_3^2}{v} \tau_i^z p^{(0)}.&
 \end{eqnarray}

The complex longitudinal dielectric permittivity of
 the deuteron subsystem of a mechanically clamped
MD$_2$XO$_4$ crystal reads
 \[
 \varepsilon_{33}^{\varepsilon'}(\omega,T) = 1 + 4\pi \chi_{33}^{\varepsilon'}
 (\omega,T), \qquad
 \varepsilon_{33}^{\varepsilon''}(\omega,T) =  4\pi \chi_{33}^{\varepsilon''}
 (\omega,T).
 \]
A numerical analysis shows that the most important contribution to the dispersion of  $\varepsilon_{33}^{\varepsilon}(\omega,T)$ is made by the first
relaxational mode $[\chi_3(1) \gg \chi_3(i)]$, while the dispersion of the complex dielectric permittivity of a mechanically clamped crystal is close to the Debye one. If the piezoelectric coupling is omitted, $\varepsilon_{33}^{\varepsilon}(\omega,T)$ transforms into the expression corresponding to~\cite{uni2009}.

\section{Comparison of the numerical results with experimental data. \\ Discussion}

Let us analyse the results of numerical calculations performed within the framework of the proposed model
for longitudinal dynamic dielectric characteristics of the  M(H$_{1-x}$D$_x)_2$XO$_4$ crystals and compare them with
the corresponding experimental data.
 It should be noted that the  theory developed in the previous sections, strictly speaking, is valid for the
MD$_2$XO$_4$ type crystals only. The experimental data are available for the  M(H$_{1-x}$D$_x)_2$XO$_4$
crystals with different deuterations $x$ $(0 \leqslant x \leqslant 1)$. The experimentally established  relaxational character
of the dielectric dispersion of $\varepsilon_{33}^*(\nu, T)$~\cite{408x,395x,361x,402x} in
these crystals, as has been already mentioned, is
associated with suppression of tunneling by  short-range
interactions. Therefore, we shall neglect the effects of proton tunneling in M(H$_{1-x}$D$_x)_2$XO$_4$.
We shall assume that the proposed  theory for these crystals is also valid if we use the averaged effective values of the model parameters
 \[
 \varepsilon(x) = \varepsilon_{\mathrm{H}}(1-x)+ \varepsilon_{\mathrm{D}}x, \qquad
 w(x) = w_{\mathrm{H}}(1-x) + w_{\mathrm{D}}x\, .
 \]

\renewcommand{\arraystretch}{0.9}
\renewcommand{\tabcolsep}{3.0pt}
\begin{table}[!h]
\caption{The obtained optimum values of the model parameters for K(H$_{1-x}$D$_x)_2$PO$_4$.}\label{tab3}
\begin{center}
\begin{tabular}{|c|c|c|c|c|c|c|c|c|c|c|c|c|c|}
\hline $x$ & $T_\mathrm{c}$ & $T_0$ & $\frac{\varepsilon}{k_\mathrm{B}}$ & $\frac{w}{k_\mathrm{B}}$ & $\frac{\nu_3(0)}{k_\mathrm{B}}$ & $\mu_{3-}, 10^{-18}$ & $\mu_{3+}, 10^{-18}$ & $\chi_{33}^0$  & $P_-$ & $R_-$ & $P_+$ & $R_+$ \\
 & (K) & (K) & (K) & (K) & (K) & (esu$\cdot$cm) & (esu$\cdot$cm) & & (s) & ({s}/{K}) & (s) & ({s}/{K})\\
\hline \hline      0.00 & 122.5 & 122.5 &  56.00 &  422.0 &  17.91 &
1.46 &   1.71 &   0.73 &   0.35 & 0.0100 &   0.43 & 0.0160 \\
   0.21 & 146.0 & 145.9 &  63.78 &  515.8 &  23.18 &   1.54 &   1.79 &   0.65 &   0.85 & 0.0095 &   1.22 & 0.0193 \\
   0.29 & 155.0 & 154.8 &  66.74 &  551.5 &  25.21 &   1.57 &   1.82 &   0.62 &   1.05 & 0.0093 &   1.51 & 0.0217 \\
   0.64 & 191.0 & 190.3 &  79.71 &  707.8 &  32.34 &   1.70 &   1.96 &   0.48 &   1.76 & 0.0385 &   2.44 & 0.0173 \\
   0.79 & 204.0 & 203.1 &  85.27 &  774.8 &  34.18 &   1.76 &   2.02 &   0.42 &   1.92 & 0.0082 &   2.65 & 0.0151 \\
   0.84 & 208.0 & 207.0 &  87.12 &  797.1 &  34.63 &   1.77 &   2.03 &   0.41 &   2.02 & 0.0081 &   2.83 & 0.0167 \\
   0.91 & 213.2 & 212.2 &  89.71 &  828.4 &  35.07 &   1.80 &   2.06 &   0.38 &   2.16 & 0.0079 &   2.88 & 0.0130 \\
   0.93 & 215.0 & 213.9 &  90.45 &  837.3 &  35.36 &   1.81 &   2.07 &   0.37 &   2.20 & 0.0079 &   3.04 & 0.0149 \\
   0.99 & 219.0 & 217.9 &  92.67 &  864.1 &  35.52 &   1.83 &   2.09 &   0.35 &   2.72 & 0.0077 &   4.21 & 0.0189 \\
   1.00 & 220.1 & 219.0 &  93.05 &  868.6 &  35.76 &   1.84 &   2.10 &   0.34 &   2.84 & 0.0077 &   4.54 & 0.0349 \\ \hline
\end{tabular}\\[2ex]
\begin{tabular}{|c|c|c|c|c|c|c|c|c|c|}
\hline $x$ & $\frac{\psi_{6}}{k_\mathrm{B}}$ & $\frac{\delta_{s6}}{k_\mathrm{B}}$ & $\frac{\delta_{a6}}{k_\mathrm{B}}$ & $\frac{\delta_{16}}{k_\mathrm{B}}$ & $c_{66}^0\cdot 10^{-10}$ & $e_{36}^0$ \\
 & (K) & (K) & (K) & (K) & (dyn/cm$^2$) & (esu/cm$^2$) \\
\hline  \hline     0.00 &--150.00 &  82.00 &--500.00 &--400.00  &   7.10
&1000.00 \\
   0.64 &--142.73 &  58.73 &--863.64 &--400.00  &   6.59  &1727.27 \\
   0.84 &--140.45 &  51.45 &--977.27 &--400.00  &   6.43  &1954.55 \\
   0.93 &--139.43 &  48.18 &--1028.41 &--400.00  &   6.36  &2056.82 \\
   1.00 &--138.64 &  45.64 &--1068.18 &--400.00  &   6.30  &2136.36 \\ \hline
\end{tabular}\\
\end{center}
\end{table}
\renewcommand{\arraystretch}{1}
\renewcommand{\tabcolsep}{1pt}

In~\cite{JPS1701}, we calculated the static longitudinal, piezoelectric, elastic, and thermal characteristics of the M(H$_{1-x}$D$_x)_2$XO$_4$ and explored their dependencies on the values of the model parameters. It was shown that at a proper choice of these values, a good quantitative agreement between the theoretical results and the corresponding experimental data was obtained. These sets of the model parameters are used herein in calculating the dynamic characteristics of  M(H$_{1-x}$D$_x)_2$XO$_4$.

The parameter $\alpha$ that  sets the time scale of the dynamic processes in  M(H$_{1-x}$D$_x)_2$XO$_4$, is determined from the condition that theoretical results for frequency dependencies of  $\varepsilon_{33}^{*}(\nu,T)$ at different temperatures agree with the experimental data. It is assumed that  $\alpha_{\mathrm{H}}$ weakly depends on temperature
  \[
  \alpha = \left(P + R |\Delta T|\right)\cdot 10^{-14}, \qquad \Delta T =
  T - T_\mathrm{c}\,.
  \]

The obtained optimum values of the model parameters are presented in table~\ref{tab3} for K(H$_{1-x}$D$_x)_2$PO$_4$ and in table~\ref{tab5} for RbH$_2$PO$_4$ and KH$_2$AsO$_4$.

\renewcommand{\arraystretch}{1.0}
\renewcommand{\tabcolsep}{3.0pt}
\begin{table}[!t]
\caption{The obtained optimum values of the model parameters for RbH$_2$PO$_4$ and KH$_2$AsO$_4$.}\label{tab5}
\begin{center}
\begin{tabular}{|c|c|c|c|c|c|c|c|c|c|}
\hline  & $T_\mathrm{c}$ & $T_0$ & $\frac{\varepsilon}{k_\mathrm{B}}$ & $\frac{w}{k_\mathrm{B}}$ & $\frac{\nu_3(0)}{k_\mathrm{B}}$ & $\mu_{3-}, 10^{-18}$ & $\mu_{3+}, 10^{-18}$ & $\chi_{33}^0$ \\
 & (K) & (K) & (K) & (K) & (K) & (esu$\cdot$cm) & (esu$\cdot$cm) &  \\
\hline \hline
RbH$_2$PO$_4$ & 147.6 & 147.6 &  60.00 &  440.0 &  29.13 &   1.50 &   2.00 &   0.40 \\
KH$_2$AsO$_4$ &  97.0 &  95.8 &  35.50 &  385.0 &  17.43 &   1.61 &   1.65 &   0.70 \\ \hline
\end{tabular}\\[2ex]
\begin{tabular}{|c|c|c|c|c|c|c|c|c|c|}
\hline  & $\frac{\psi_{6}}{k_\mathrm{B}}$ & $\frac{\delta_{s6}}{k_\mathrm{B}}$ & $\frac{\delta_{a6}}{k_\mathrm{B}}$ & $\frac{\delta_{16}}{k_\mathrm{B}}$ & $c_{66}^0\cdot 10^{-10}$ & $e_{36}^0$ \\
 & (K) & (K) & (K) & (K) & (dyn/cm$^2$) & (esu/cm$^2$) \\
\hline\hline
RbH$_2$PO$_4$ &--130.00 &  50.00 &--500.00 &--300.00  &   5.90   &3000.00 \\
KH$_2$AsO$_4$ & --170.00 & 130.00 &--500.00 &--500.00  &   7.50  &3000.00 \\ \hline
\end{tabular}\\[2ex]
\begin{tabular}{|c|c|c|c|c|c|c|c|c|c|}
\hline  & $P_-$ & $R_-$ & $P_+$ & $R_+$ \\
 & ($s$) & ($\frac{s}{K}$) & ($s$) & ($\frac{s}{K}$) \\
\hline\hline
RbH$_{2}$PO$_{4}$ &   0.55 & 0.0080 &   0.93 & 0.0140 \\
KH$_2$AsO$_4$ &   0.47 & 0.0160 &   0.61 & 0.0190 \\ \hline
\end{tabular}
\end{center}
\end{table}
\renewcommand{\arraystretch}{1}
\renewcommand{\tabcolsep}{1pt}

Note that $ \mu_{3+},P_{+},R_{+}$ and  $\mu_{3-}, P_{-}R_{-}$ correspond to the paraelectric and ferroelectric phases, respectively.
\begin{figure}[!h]
\centerline{
 \includegraphics[scale=0.57]{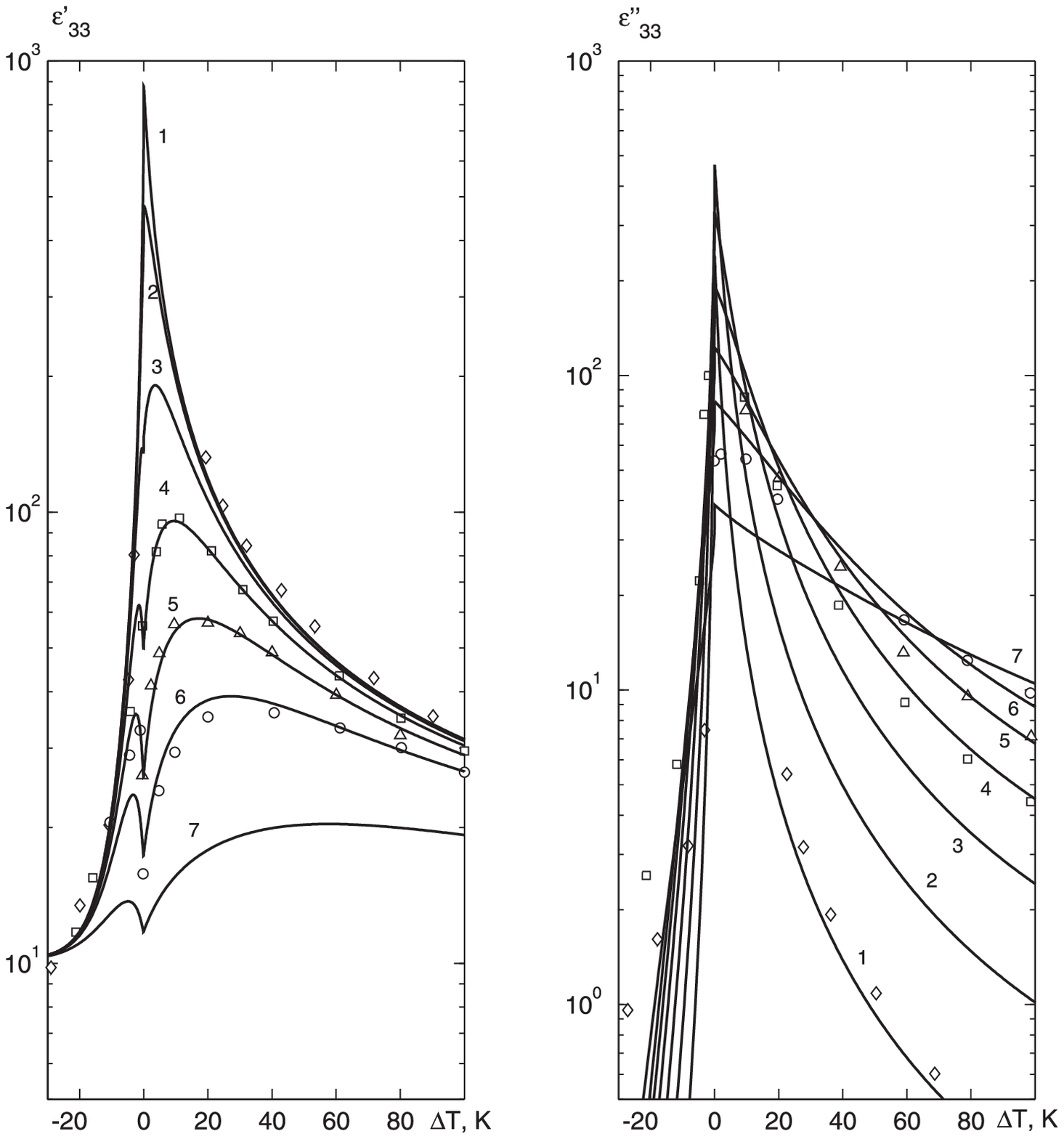}
}
\caption[]{The temperature dependence of \ $\varepsilon'_{33}$ and \
$\varepsilon''_{33}$ in  KH$_{2}$PO$_{4}$ at different frequencies $\nu$ (GHz): 9.2~-- 1, \ra{\e{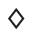}} \cite{395x}; 33.2~-- 2; 80~-- 3;
154.2~-- 4, \ra{\e{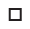}} \cite{408x}; 249~-- 5,
\ra{\e{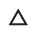}} \cite{408x}; 372~-- 6, \ra{\e{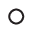}} \cite{408x};
800~-- 7.  Symbols are experimental points; lines are the theoretical values.} \label{e33reim_log}
\end{figure}
\begin{figure}[!h]
\begin{center}
 \includegraphics[scale=0.52]{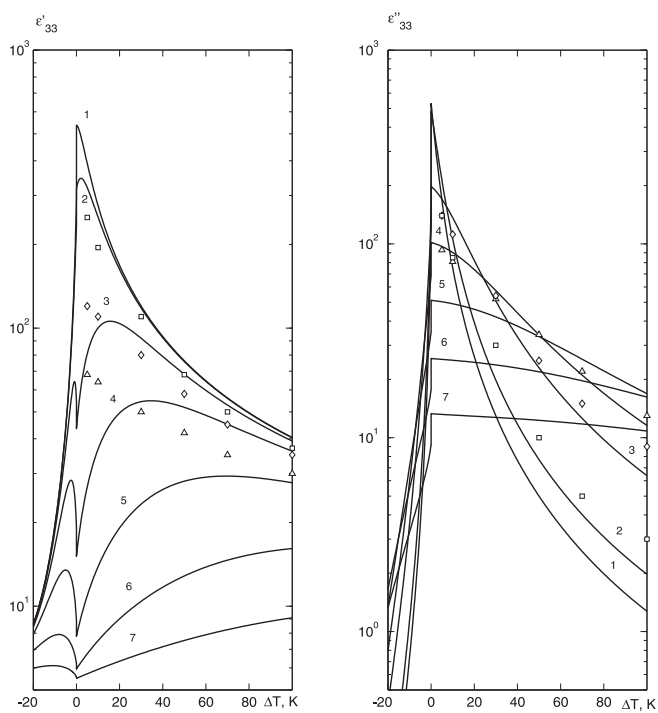}
\end{center}
\vspace{-0.4cm}
\caption[]{The temperature dependence of  $\varepsilon'_{33}$ and
$\varepsilon''_{33}$ in  KD$_{2}$PO$_{4}$ at different frequencies $\nu$ (GHz):
1.93~-- 1; 3.0~-- 2, \ra{\e{s0.eps}} \cite{361x}; 10.0~-- 3,
\ra{\e{d0.eps}} \cite{361x}; 20.0~-- 4, \ra{\e{up0.eps}} \cite{361x};
40.0~-- 5; 80.0~-- 6; 154.2~-- 7. Symbols are experimental points; lines are the theoretical values.} \label{e33reim_x1_log}
\end{figure}
\begin{figure}[!h]
\begin{center}
 \includegraphics[scale=0.52]{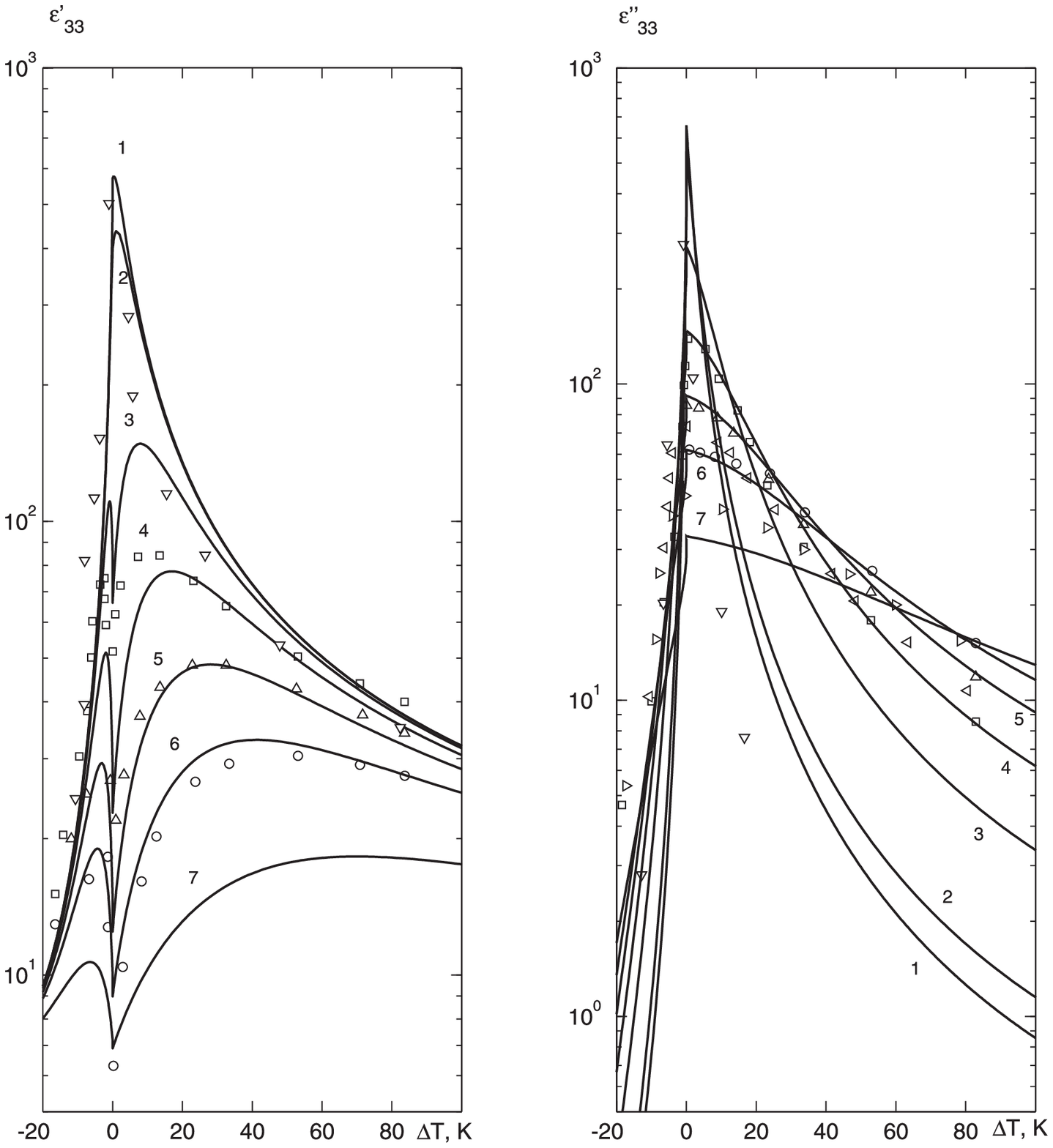}
\end{center}
\vspace{-0.4cm}
\caption[]{ The temperature dependence of  $\varepsilon'_{33}$ and
$\varepsilon''_{33}$  in  RbH$_{2}$PO$_{4}$ at different frequencies $\nu$ (GHz): 0.25~-- 1; 10.0~-- 2; 27.0~-- 3,
\ra{\e{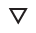}} \cite{403x}; 154.2~-- 4, \ra{\e{s0.eps}} \cite{410x};
250.2~-- 5, \raisebox{-0.0ex}[0cm][0cm]{\e{up0.eps}} \cite{410x}; 372.0~-- 6,
\ra{\e{o0.eps}} \cite{410x}; 700.0~-- 7.  Symbols are experimental points; lines are the theoretical values.} \label{e33reim_RDP_log}
\end{figure}
\begin{figure}[!h]
\begin{center}
 \includegraphics[scale=0.52]{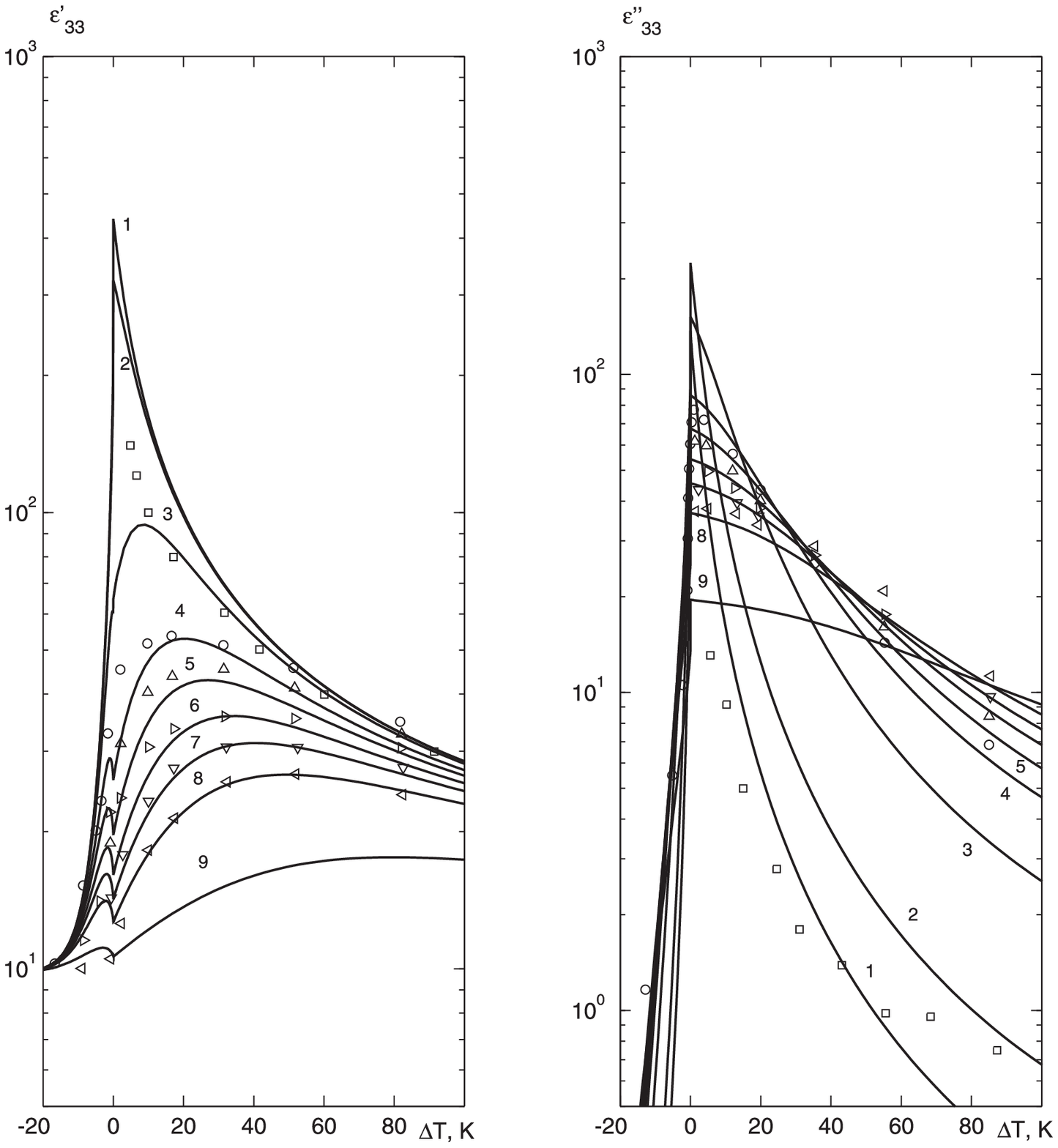}
\end{center}
\vspace{-0.4cm}
\caption[]{The temperature dependence of \ $\varepsilon'_{33}$ and \
$\varepsilon''_{33}$ in  KH$_{2}$AsO$_{4}$ at different frequencies $\nu$ (GHz): 9.2~-- 1, \!\ra{\e{s0.eps}} \cite{395x}; 20.8~-- 2; 80.0~-- 3;
154.2~-- 4, \raisebox{-0.0ex}{\e{o0.eps}} \cite{410x}; 198.9~-- 5,
\ra{\e{up0.eps}} \cite{410x}; 250.2~-- 6,
\ra{\e{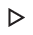}} \cite{410x}; 7, \ra{\e{do0.eps}} \cite{410x}; 372.0~-- 8, \ra{\e{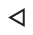}} \cite{410x}; 700.0~-- 9.  Symbols are experimental points; lines are the theoretical values.}
\label{e33reim_KDA_log}
\end{figure}
The temperature dependencies of the real and imaginary parts of the permittivity
 $\varepsilon_{33}'(\nu,T)$ and
 $\varepsilon_{33}''(\nu,T)$ at different frequencies for the  KH$_2$PO$_4$, KD$_2$PO$_4$,
 RbH$_2$PO$_4$, and KH$_2$AsO$_4$ crystals are shown in figures~\ref{e33reim_log}--\ref{e33reim_KDA_log}.
Starting from a certain frequency $\nu_k$, the low-frequency maximum in the temperature curve of  $\varepsilon_{33}'(\nu,T)$
is replaced with a sharp minimum at  $\Delta T = 0$~K which widens and deepens with an increasing frequency that reaches $\varepsilon_{33}^{0}$ at  $\nu \sim 10^{12}$~Hz. In KH$_2$PO$_4$ $\nu_k = 33.2$~GHz,
in KD$_2$PO$_4$ $\nu_k = 1.4$~GHz, in RbH$_2$PO$_4$ $\nu_k = 20.5$~GHz,
in KH$_2$AsO$_4$  $\nu_k = 20.8$~GHz. The maximum of  $\varepsilon_{33}'(\nu,T)$ at $\Delta T_n = |T_n - T_\mathrm{c}|$ decreases and smears out with an increasing frequency, whereas the magnitude of  $\Delta T_n$ increases. With increasing frequency the magnitude of
 $\varepsilon_{33}'(\nu)$ decreases at all $\Delta T = |T -
 T_\mathrm{c}|$. The maximal values of
 $\varepsilon_{33}'(\nu)$ as well as the values of  $\Delta T_n$ are much larger in the paraelectric phase than in the ferroelectric phase.
 The dispersion of a real part of the permittivity
 $\varepsilon_{33}'(\nu,T)$ in the ferrroelectric phase
 is observed in a narrow temperature range $\Delta T \sim
 20$~K, whereas in the paraelectric phase, $\Delta T$ is much larger, being of the order of  200~K.

Let us note that taking into account the piezoelectric coupling, the calculated minimal values of  $\varepsilon_{33}'(\nu)$ at $\Delta T
 =0$  at different frequencies  are larger than those obtained  within the model without the piezoelectric coupling.

At a decreasing $\Delta T$ in the ferroelectric phase, the value of
 $\varepsilon''_{33}(\nu)$ increases, has a maximum at  $\Delta T
 =0$, and decreases with an increasing $\Delta T$ in the paraelectric phase.
  At an increasing frequency, the maximal value of
 $\varepsilon''_{33}(\nu)$ and the rate of its change with an increasing  $\Delta
 T$ diminish.

At   $\nu_k$, the values of  $\varepsilon_{33}'(\nu,T) =
 \varepsilon_{33}''(\nu,T)$ are
465 in KH$_2$PO$_4$, 520 in  KD$_2$PO$_4$,
562 in RbH$_2$PO$_4$, and   330 in KH$_2$AsO$_4$.

The proposed theory provides a good quantitative agreement with the experiment for  KH$_2$PO$_4$ (figure~\ref{e33reim_log}) and a little worse agreement for the data of~\cite{361x} for  KD$_2$PO$_4$ (figure~\ref{e33reim_x1_log}), especially at $\Delta T <
 20$~K for $\varepsilon'_{33}(\nu,T)$. However, it should, be noted that the  values of $\varepsilon'_{33}(\nu,T)$ obtained in~\cite{361x} at frequencies above  1~GHz have maxima at $\Delta T = 0$~K, rather than  minima.

The temperature dependence of  $\varepsilon_{33}^{*}(\nu,T)$ in RbH$_2$PO$_4$ measured in~\cite{410x} is appropriately and  well described by the present theory, except for the values of $\varepsilon_{33}'(\nu,T)$ at $\nu =
 154.2$~GHz and  $\Delta T < 20$~K (figure~\ref{e33reim_RDP_log}). The theory and experimental data of~\cite{403x} for $\varepsilon_{33}^{*}(\nu,T)$
at $\nu = 27$~GHz are also in a good agreement. The obtained theoretical results for  $\varepsilon_{33}^{*}(\nu,T)$ at
 198 and 366 are only in qualitative agreement with the data of~\cite{397x}, which, in their turn, are in disagreement with the results of other measurements of~\cite{410x}.

The calculated temperature dependencies of
 $\varepsilon_{33}^{\varepsilon'}(\nu,T)$ and
 $\varepsilon_{33}^{\varepsilon''}(\nu,T)$ accord well with the ones measured in~\cite{410x} for KH$_2$AsO$_4$ at different frequencies starting  from the submillimeter range (figure~\ref{e33reim_KDA_log}). The data for $\varepsilon_{33}^{*}(\nu,T)$ obtained in~\cite{395x} at $\nu = 9.2$~GHz are in a somewhat worse agreement with the theory, especially at $\Delta T < 20$~K.

\begin{figure}[!h]
\begin{center}
 \includegraphics[width=0.74\textwidth]{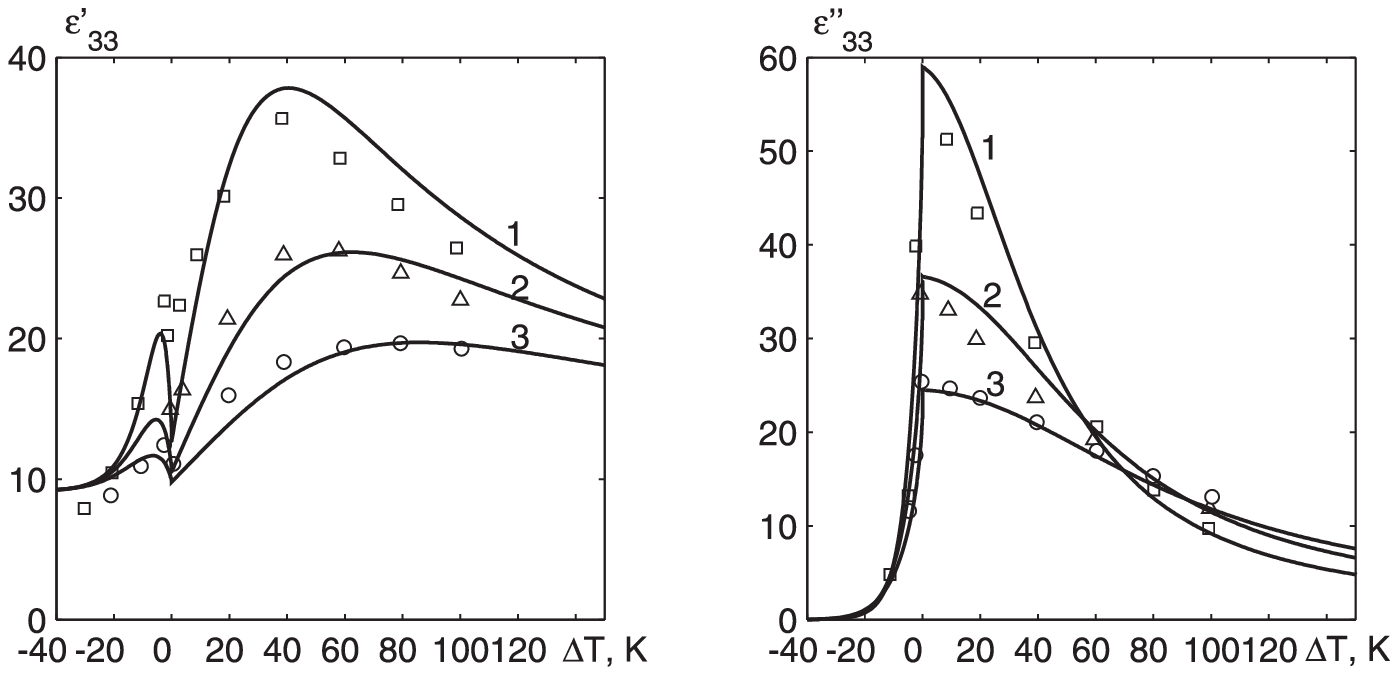} 
\end{center}
\caption[]{The temperature dependence of \ $\varepsilon'_{33}$ and
$\varepsilon''_{33}$ in K(H$_{0.79}$D$_{0.21}$)$_{2}$PO$_{4}$ at different frequencies $\nu$~(GHz)~\cite{408x}: 154.2~-- 1,
\ra{\e{s0.eps}}; \ 249.0~-- 2, \ra{\e{up0.eps}}; \ 372.0~-- 3,
\ra{\e{o0.eps}}. Symbols are experimental points; lines are the theoretical values.} \label{e33Treimx021}
\end{figure}

\begin{figure}[!h]
\begin{center}
 \includegraphics[width=0.74\textwidth]{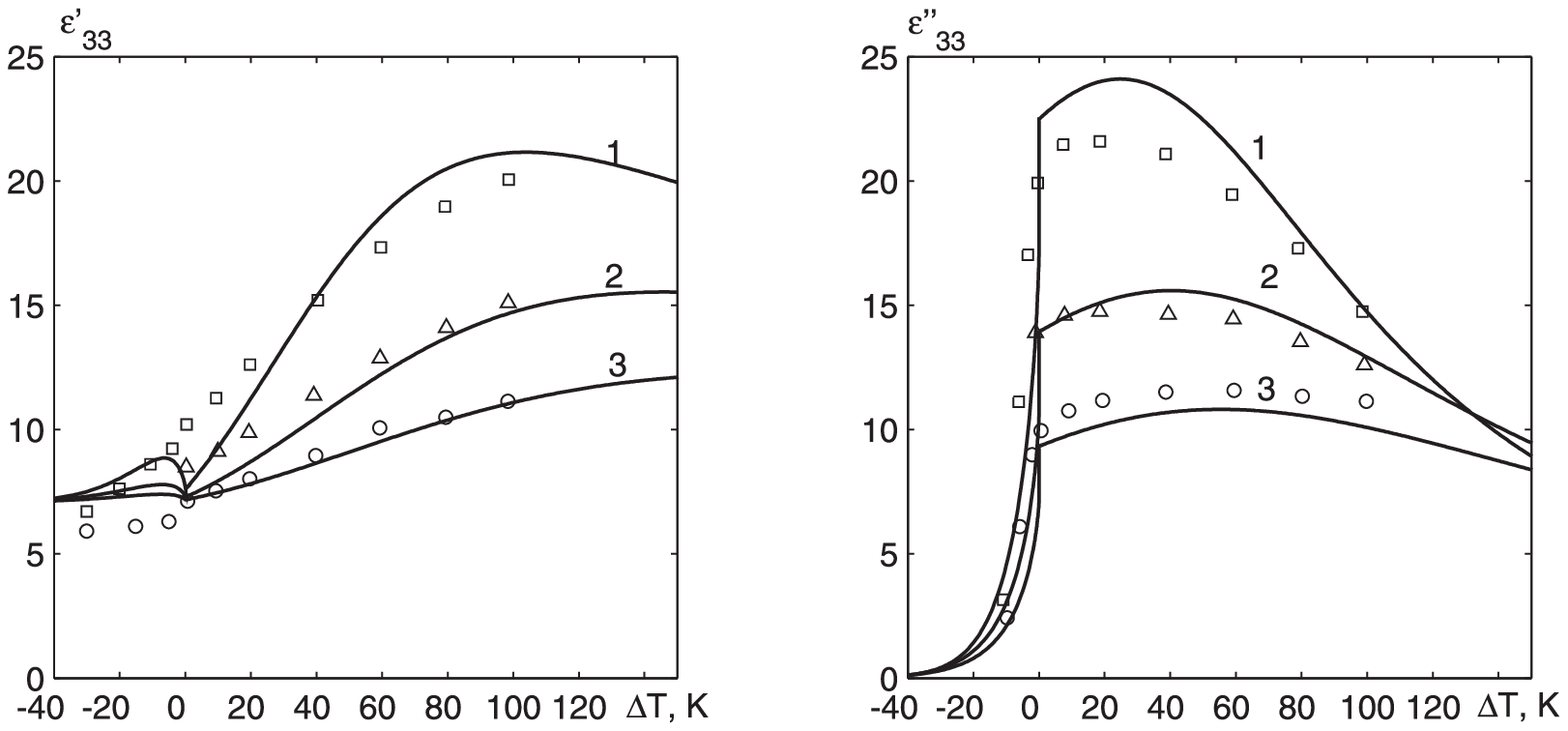}
\end{center}
\caption[]{The temperature dependence of \ $\varepsilon'_{33}$ and
$\varepsilon''_{33}$ in K(H$_{0.36}$D$_{0.64}$)$_{2}$PO$_{4}$  at different frequencies $\nu$~(GHz)~\cite{408x}: 154.2~-- 1,
\ra{\e{s0.eps}}; \ 249.0~-- 2, \ra{\e{up0.eps}}; \ 372.0~-- 3,
\ra{\e{o0.eps}}. Symbols are experimental points; lines are the theoretical values.} \label{e33Treimx064}
\end{figure}

\begin{figure}[!h]
\begin{center}
 \includegraphics[width=0.74\textwidth]{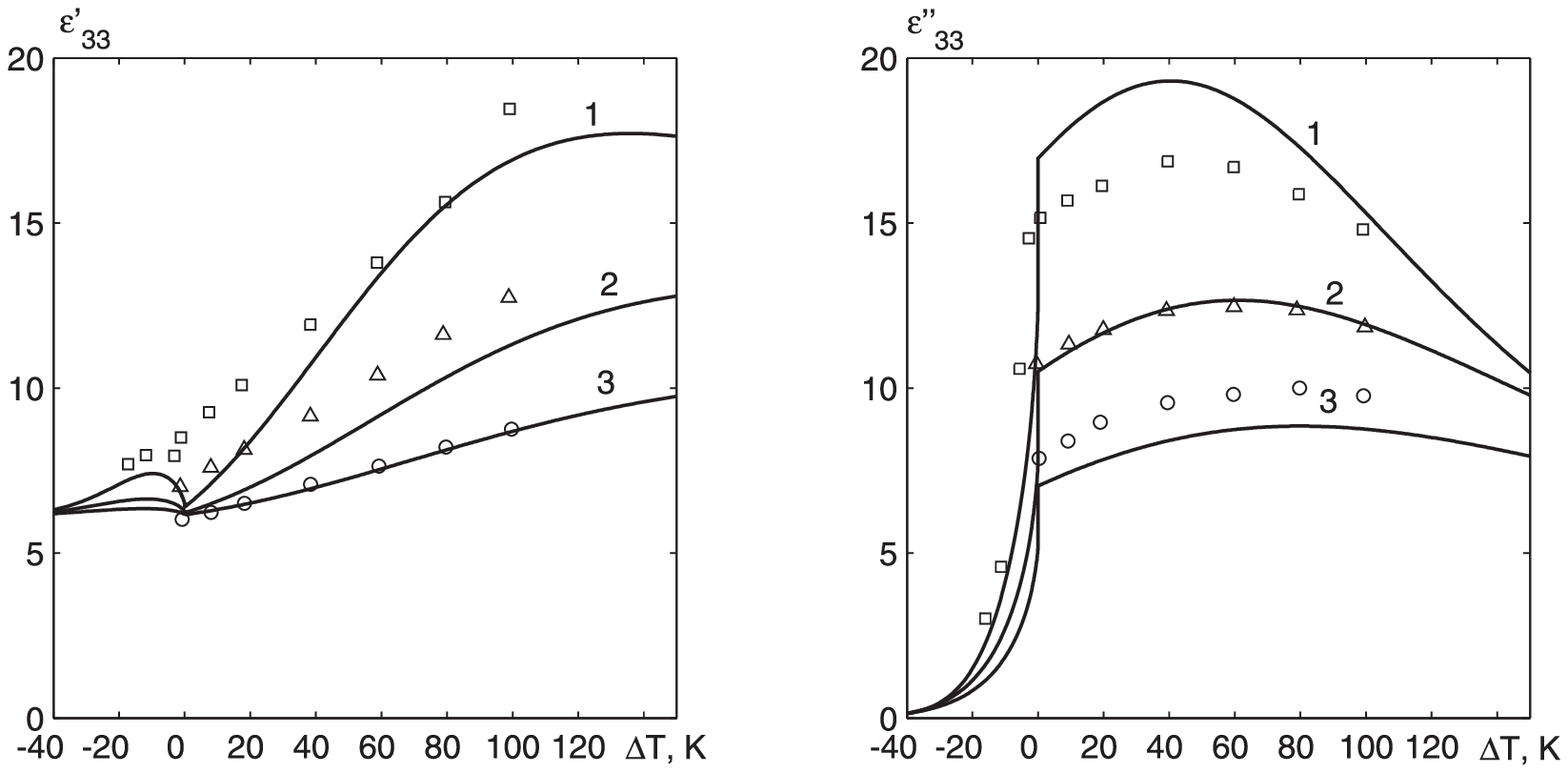}
\end{center}
\vspace{-3mm}
\caption[]{The temperature dependence of \ $\varepsilon'_{33}$ and
$\varepsilon''_{33}$ in K(H$_{0.16}$D$_{0.84}$)$_{2}$PO$_{4}$  at different frequencies $\nu$~(GHz)~\cite{408x}: 154.2~-- 1,
\ra{\e{s0.eps}}; \ 249.0~-- 2, \ra{\e{up0.eps}}; \ 372.0~-- 3,
\ra{\e{o0.eps}}. Symbols are experimental points; lines are the theoretical values.} \label{e33Treimx084}
\end{figure}

\begin{figure}[!h]
\begin{center}
 \includegraphics[width=0.74\textwidth]{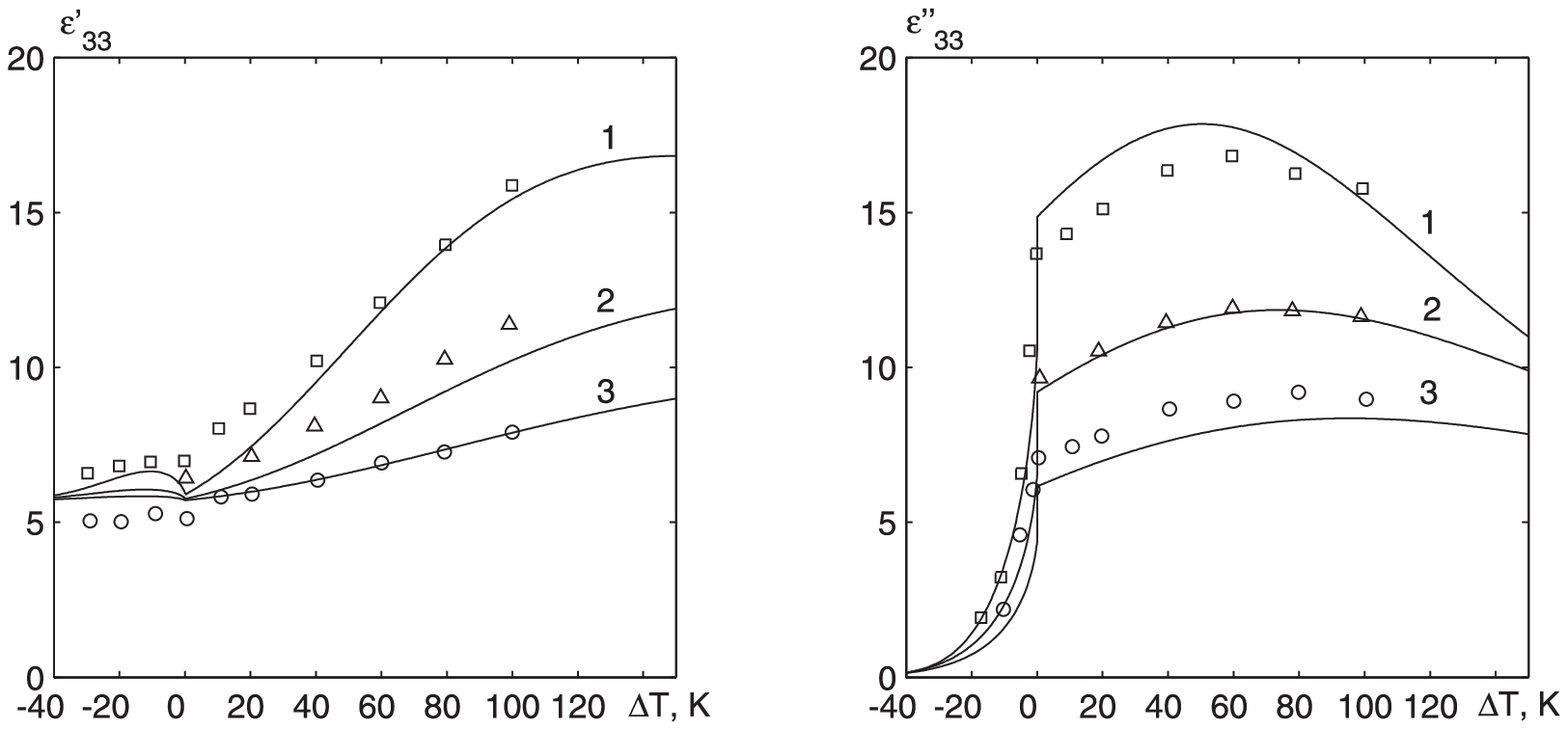}
\end{center}
\caption[]{The temperature dependence of \ $\varepsilon'_{33}$ and
$\varepsilon''_{33}$ in K(H$_{0.07}$D$_{0.93}$)$_{2}$PO$_{4}$  at different frequencies $\nu$~(GHz)~\cite{408x}: 154.2~-- 1,
\ra{\e{s0.eps}}; \ 249.0~-- 2, \ra{\e{up0.eps}}; \ 372.0~-- 3,
\ra{\e{o0.eps}}. Symbols are experimental points; lines are the theoretical values.} \label{e33Treimx093}
\end{figure}

\begin{figure}[!h]
\begin{center}
 \includegraphics[width=0.74\textwidth]{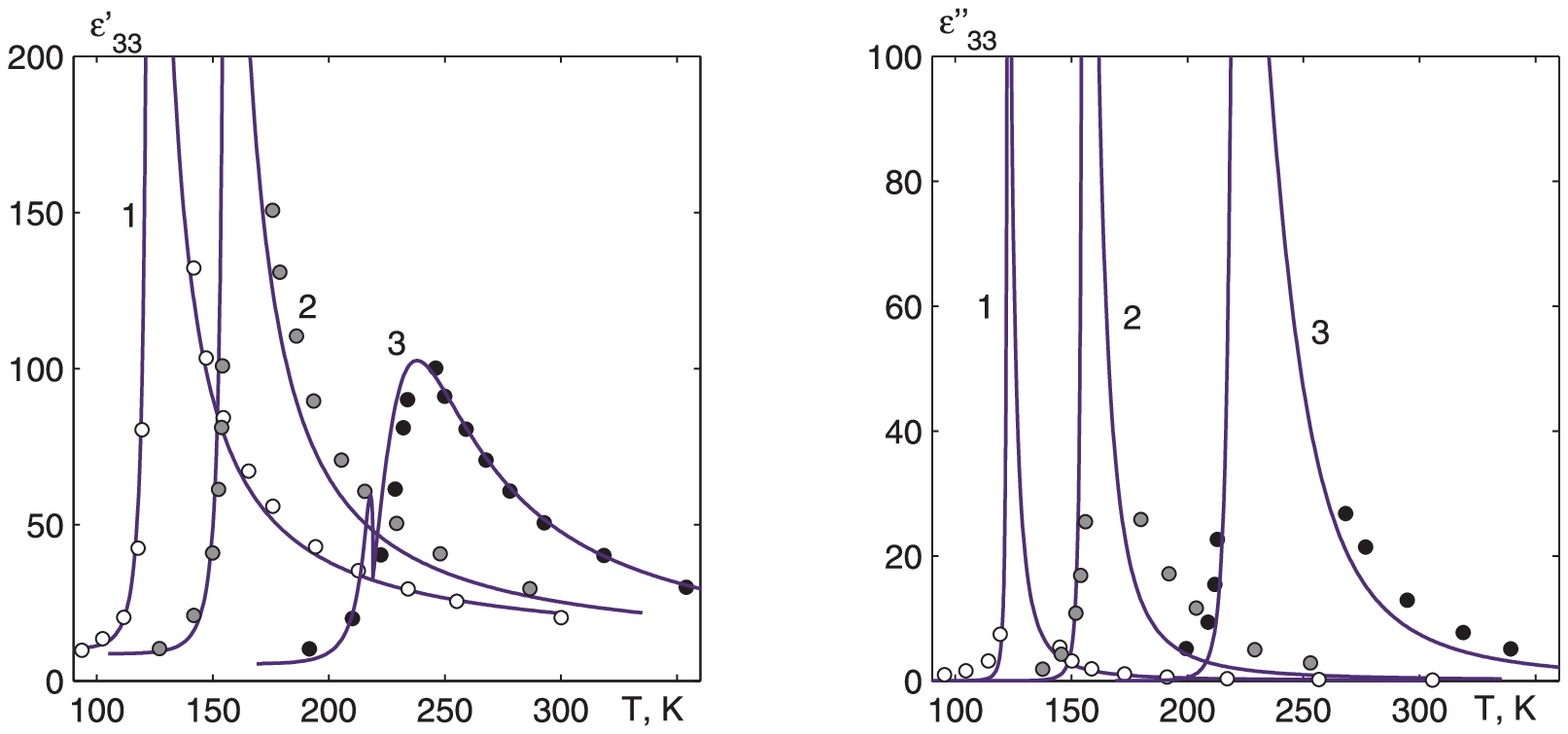}
\end{center}
\caption[]{ The  temperature  dependence of  $\varepsilon'_{33}$ and
$\varepsilon''_{33}$ in K(H$_{1-x}$D$_{x}$)$_{2}$PO$_{4}$ at
$\nu$=9.2~GHz and for \linebreak different $x$~\cite{395x}:  0.0~-- 1, \ra{\e{o0.eps}};
0.29~-- 2, \ra{\e{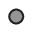}};  0.99~-- 3,  \ra{\e{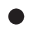}}. Symbols are experimental points; lines are the theoretical values.}
\label{e33reimx0_29_99}
\end{figure}

\begin{figure}[!h]
\begin{center}
 \includegraphics[width=0.74\textwidth]{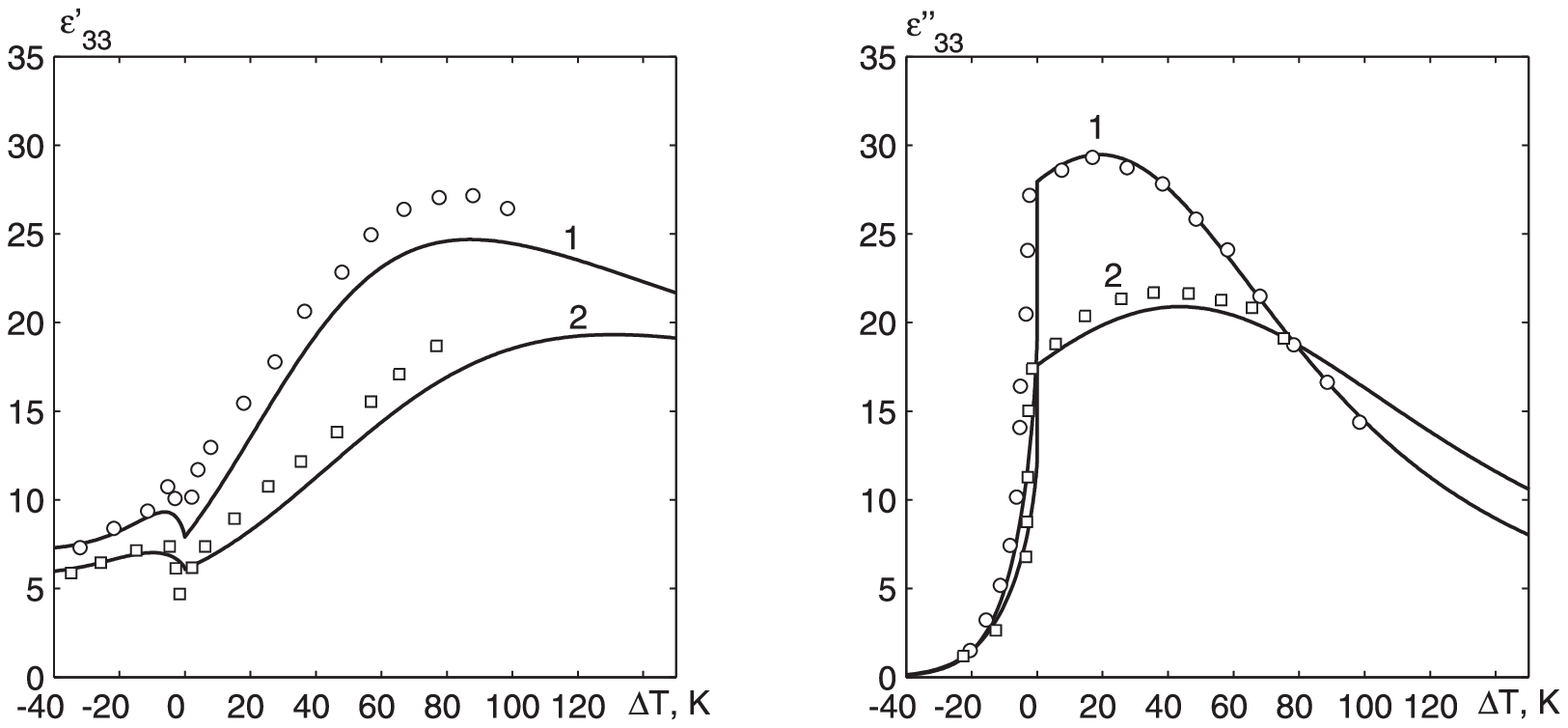}
\end{center}
\vspace{-3mm}
\caption[]{The temperature dependence of  $\varepsilon'_{33}$ and
$\varepsilon''_{33}$ in K(H$_{1-x}$D$_{x}$)$_{2}$PO$_{4}$ at
$\nu$=138.6~GHz and for different $x$~\cite{399x}: 0.63~-- 1,
\ra{\e{o0.eps}}; 0.91~-- 2, \ra{\e{s0.eps}}. Symbols are experimental points; lines are the theoretical values.}
\label{e33reimx063_091}
\end{figure}

\begin{figure}[!h]
\begin{center}
 \includegraphics[width=0.74\textwidth]{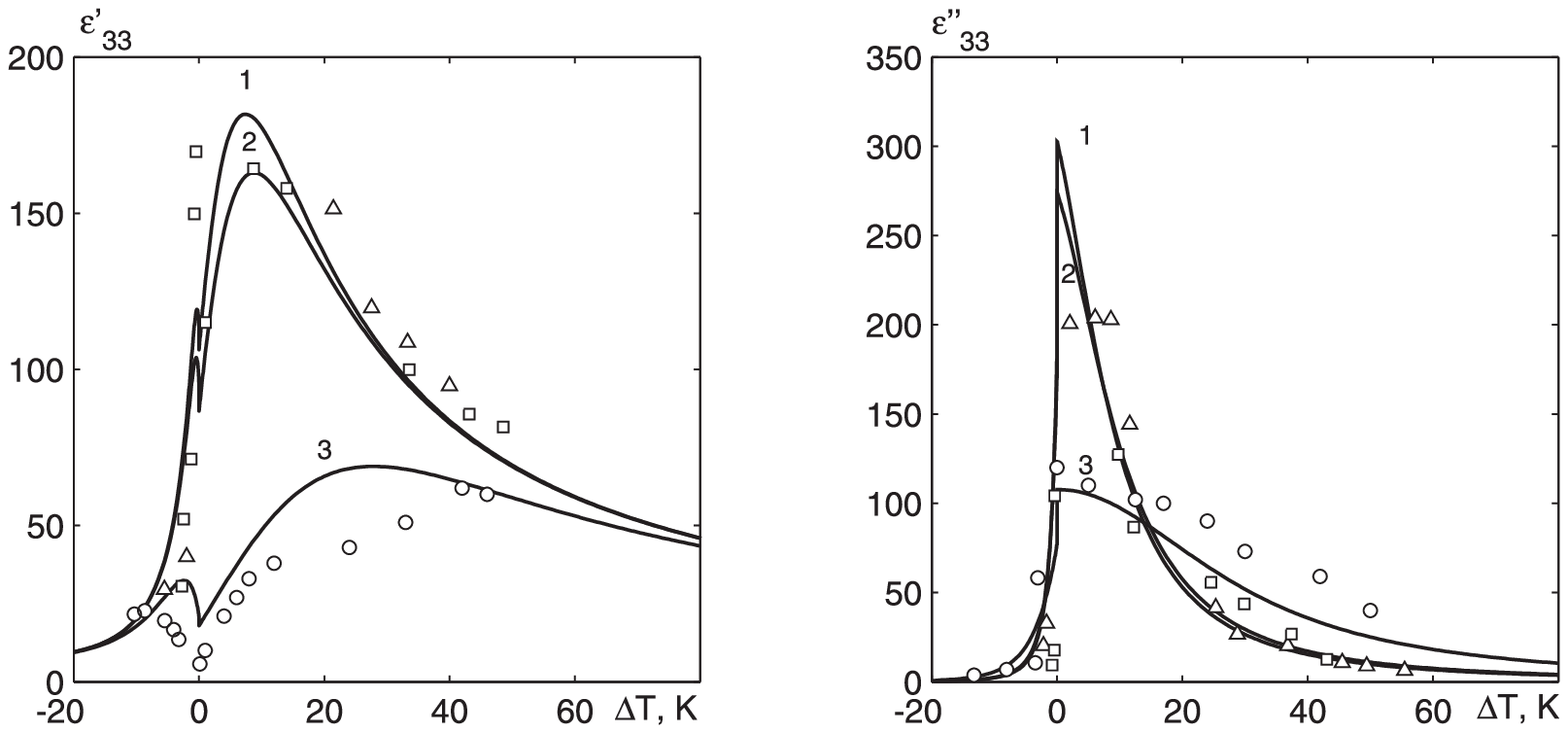}
\end{center}
\vspace{-3mm}
\caption[]{The temperature dependence of  $\varepsilon'_{33}$ and
\ $\varepsilon''_{33}$ in K(H$_{0.22}$D$_{0.78}$)$_{2}$PO$_{4}$ at different frequencies $\nu$~(GHz)~\cite{402x}: 8.6~-- 1,
\ra{\e{up0.eps}}; 9.7~-- 2, \ra{\e{s0.eps}}; 26.5~-- 3,
\ra{\e{o0.eps}}. Symbols are experimental points; lines are the theoretical values.} \label{e33reimx078}
\end{figure}

\begin{figure}[!h]
\begin{center}
 \includegraphics[width=0.74\textwidth]{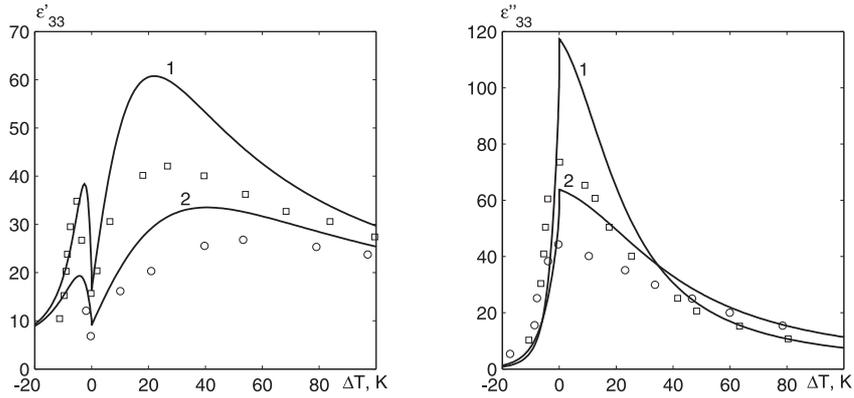}
\end{center}
\vspace{-3mm}
\caption[]{The temperature dependence of  $\varepsilon'_{33}$ and
$\varepsilon''_{33}$ in RbH$_{2}$PO$_{4}$ at different frequencies $\nu$~(GHz)~\cite{397x}:
198.0~-- 1, \ra{\e{s0.eps}}; 366.0~-- 2,
\ra{\e{o0.eps}}. Symbols are experimental points; lines are the theoretical values.} \label{e33reim_RDP_36G}
\end{figure}

Figures~\ref{e33Treimx021}--\ref{e33reim_RDP_36G} contain the calculated temperature dependencies of the
real and imaginary parts of longitudinal dynamic dielectric permittivity $\varepsilon'_{33}(\nu,T)$ and
 $\varepsilon''_{33}(\nu,T)$
of clamped K(H$_{1-x}$D$_x)_2$PO$_4$ crystals at different deuterations $x$ and frequencies along with the
corresponding experimental data.

With an increasing deuteration $x$ in
 K(H$_{1-x}$D$_x)_2$PO$_4$, the magnitude of
 $\varepsilon_{33}'(\nu,T)$ decreases, whereas $\Delta T_n$ increases.

At an isomorphic replacement K $\to$ Rb, P$\to$ As, the maximal values of  $\varepsilon_{33}'(\nu,T)$ remain almost unchanged, whereas $\Delta T_n$ slightly increase.

It should be noted that in the MH$_2$XO$_4$,  the experimental data of~\cite{352x,395x,398x,410x} cor\-res\-pond to the region of
dielectric permittivity dispersion. At the same time, for KD$_2$PO$_4$ in the measurements of~\cite{408x}, the submillimeter frequencies correspond to a high-frequency ``tail'' of the dispersion, whereas in the data of~\cite{361x}, this is the low-frequency tail.
Further experimental measurements of  $\varepsilon_{33}^{*}(\nu,T)$ at $\nu>10$ are required to
evaluate the validity of the calculated  $\varepsilon_{33}^{*}(\nu,T)$.

\begin{figure}[!h]
\vspace{-1mm}
\begin{center}
 \includegraphics[width=0.7\textwidth]{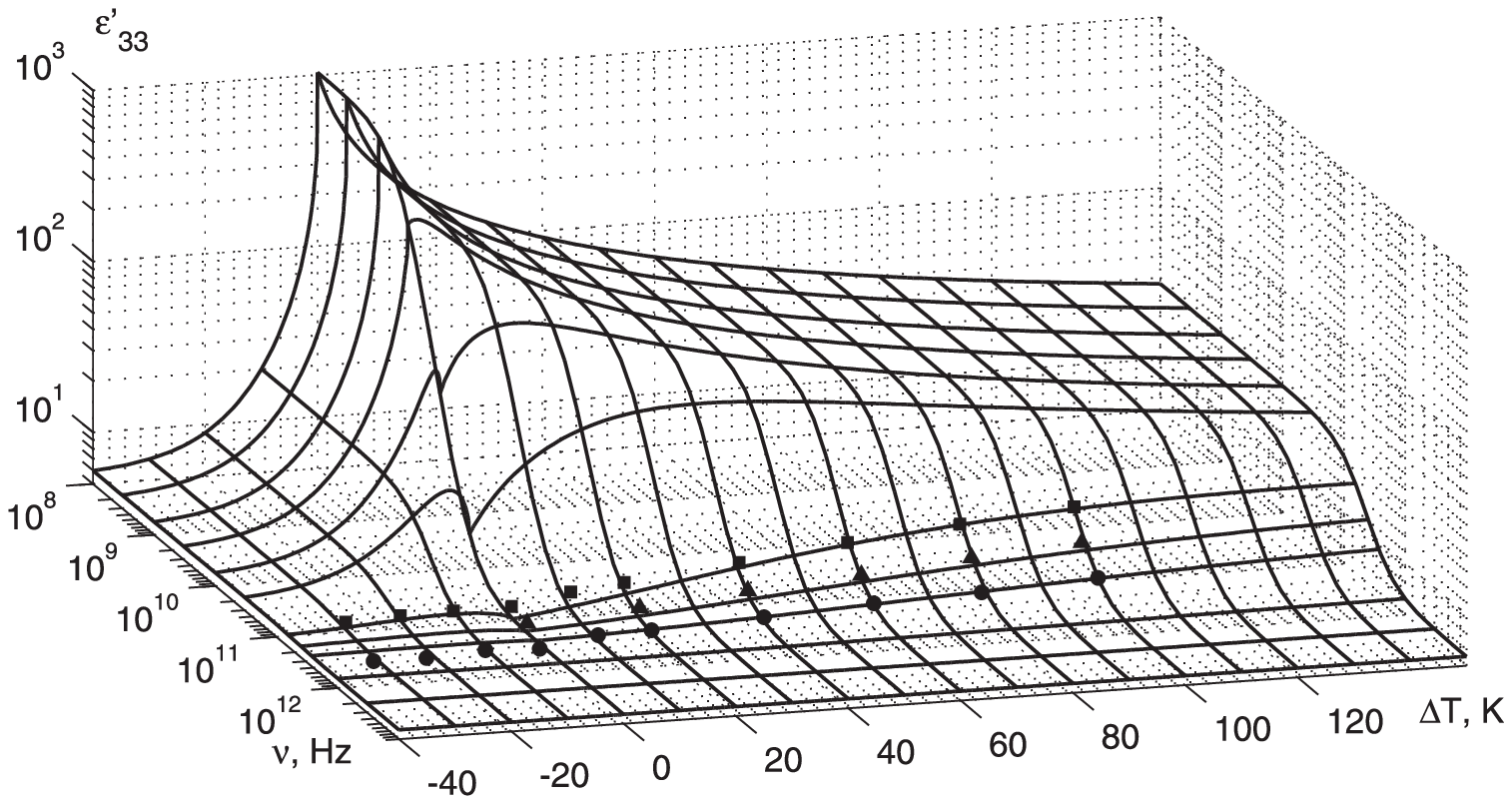}
\end{center}
\vspace{-4mm}
\caption[]{The frequency-temperature dependence of $\varepsilon'_{33}$ in
K(H$_{0.07}$D$_{0.93}$)$_{2}$PO$_{4}$. \ra{\e{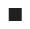}},
\ra{\e{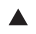}}, \ra{\e{o1.eps}}~--~\cite{408x}. Symbols are experimental points; lines are the theoretical values.}
\label{e33nuT_x093}
\end{figure}
\begin{figure}[!h]
\vspace{-3mm}
\begin{center}
 \includegraphics[width=0.7\textwidth]{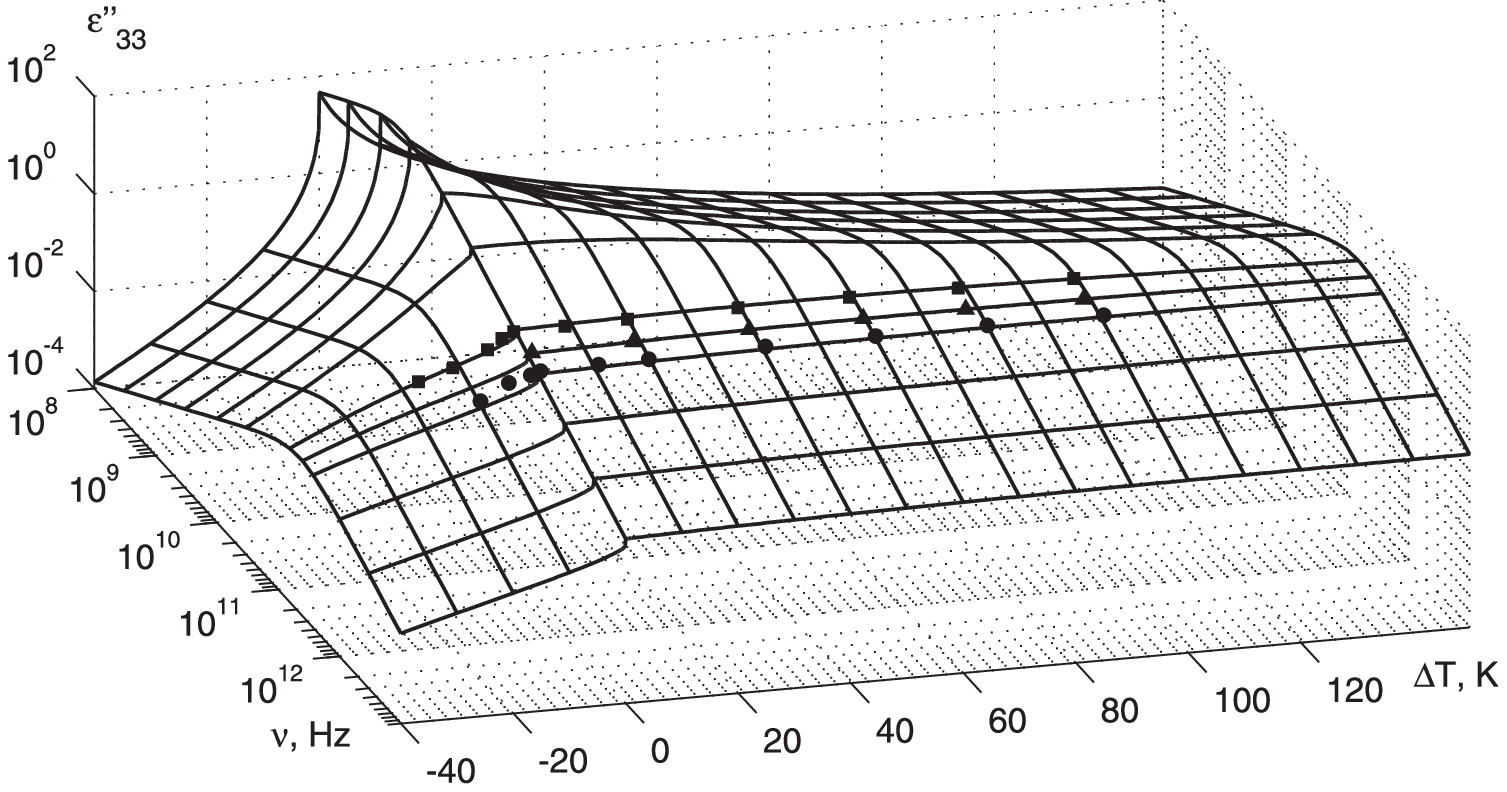}
\end{center}
\vspace{-4mm}
\caption[]{The frequency-temperature dependence of  $\varepsilon''_{33}$ in
K(H$_{0.07}$D$_{0.93}$)$_{2}$PO$_{4}$. \ra{\e{s1.eps}},
\ra{\e{up1.eps}}, \ra{\e{o1.eps}}~--~\cite{408x}. Symbols are experimental points; lines are the theoretical values.}
\label{e33nuTim_x093}
\end{figure}
\begin{figure}[!h]
\vspace{-3mm}
\begin{center}
 \includegraphics[width=0.7\textwidth]{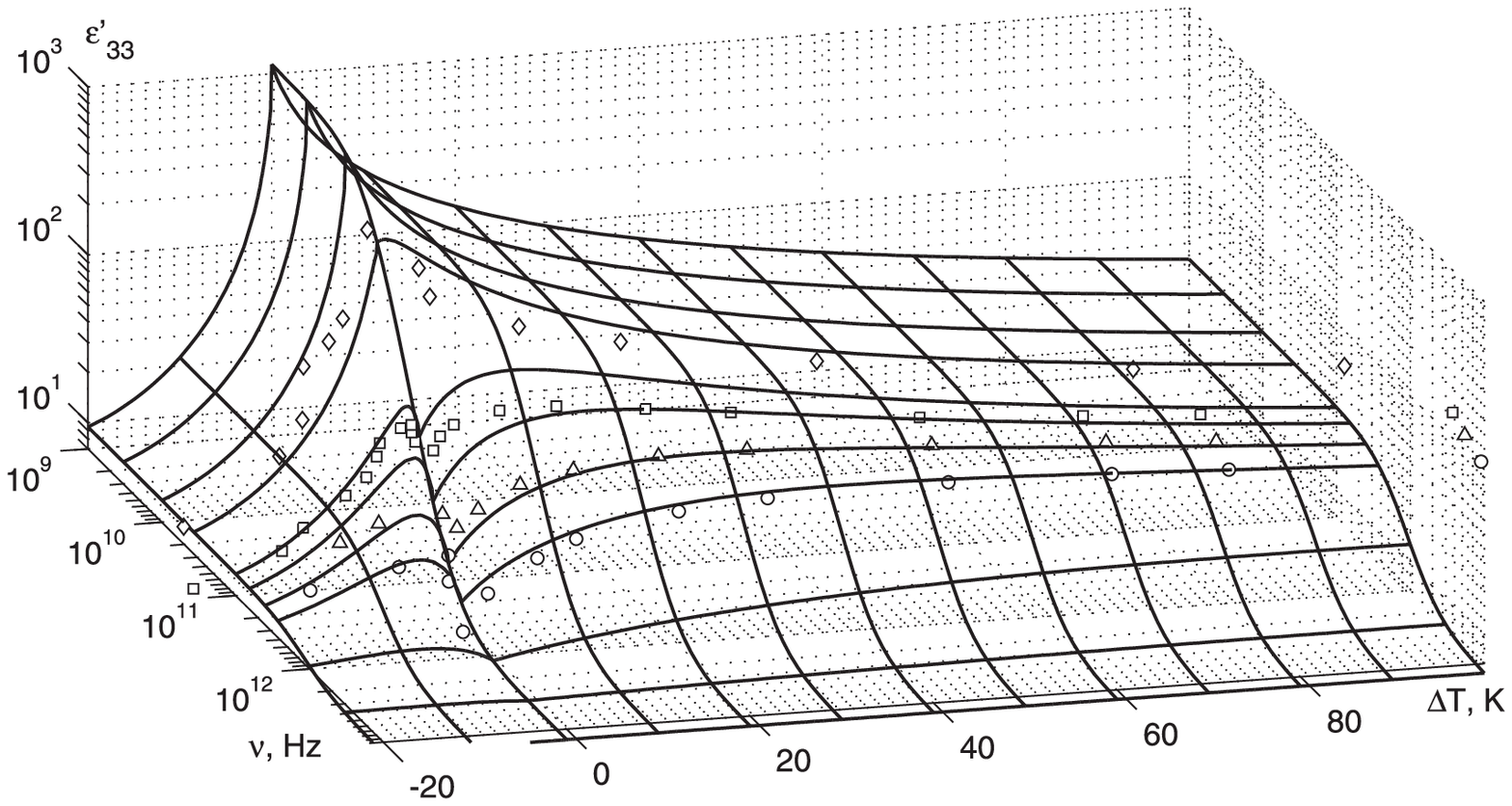}
\end{center}
\vspace{-4mm}
\caption[]{The frequency-temperature dependence of $\varepsilon'_{33}$ in
RbH$_{2}$PO$_{4}$. \ra{\e{d0.eps}}~--~\cite{403x}; \ra{\e{s0.eps}},
\ra{\e{up0.eps}}, \ra{\e{o0.eps}}~--~\cite{410x}. Symbols are experimental points; lines are the theoretical values.}
\label{e33nuT_RDP}
\end{figure}
\begin{figure}[!h]
\vspace{1mm}
\begin{center}
 \includegraphics[width=0.7\textwidth]{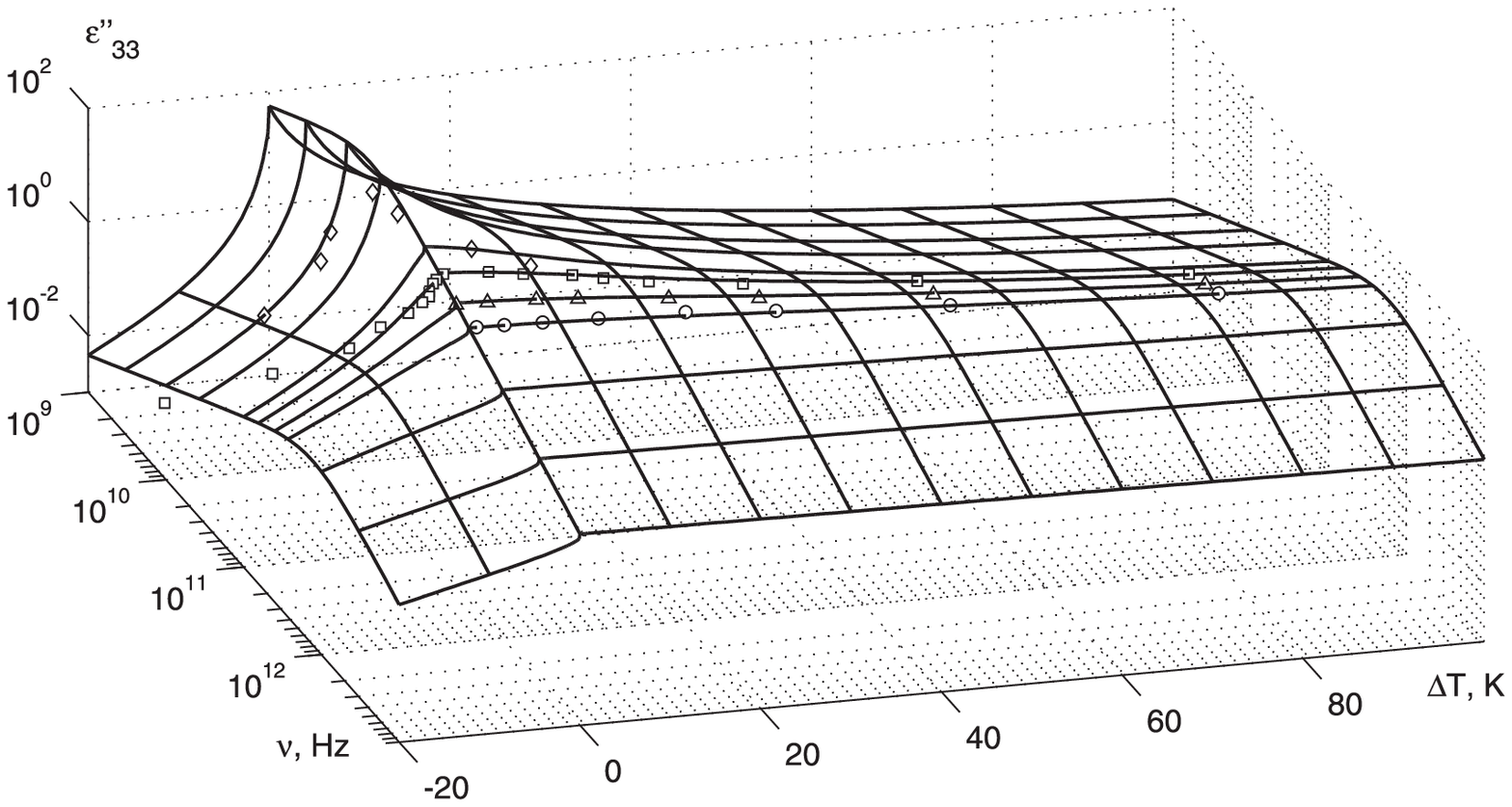}
\end{center}
\caption[]{The frequency-temperature dependence of $\varepsilon''_{33}$ in
RbH$_{2}$PO$_{4}$. \ra{\e{d0.eps}}~--~\cite{403x}; \ra{\e{s0.eps}},
\ra{\e{up0.eps}}, \ra{\e{o0.eps}}~--~\cite{410x}. Symbols are experimental points; lines are the theoretical values.}
\label{e33nuTim_RDP}
\end{figure}
\begin{figure}[!h]
\begin{center}
 \includegraphics[width=0.7\textwidth]{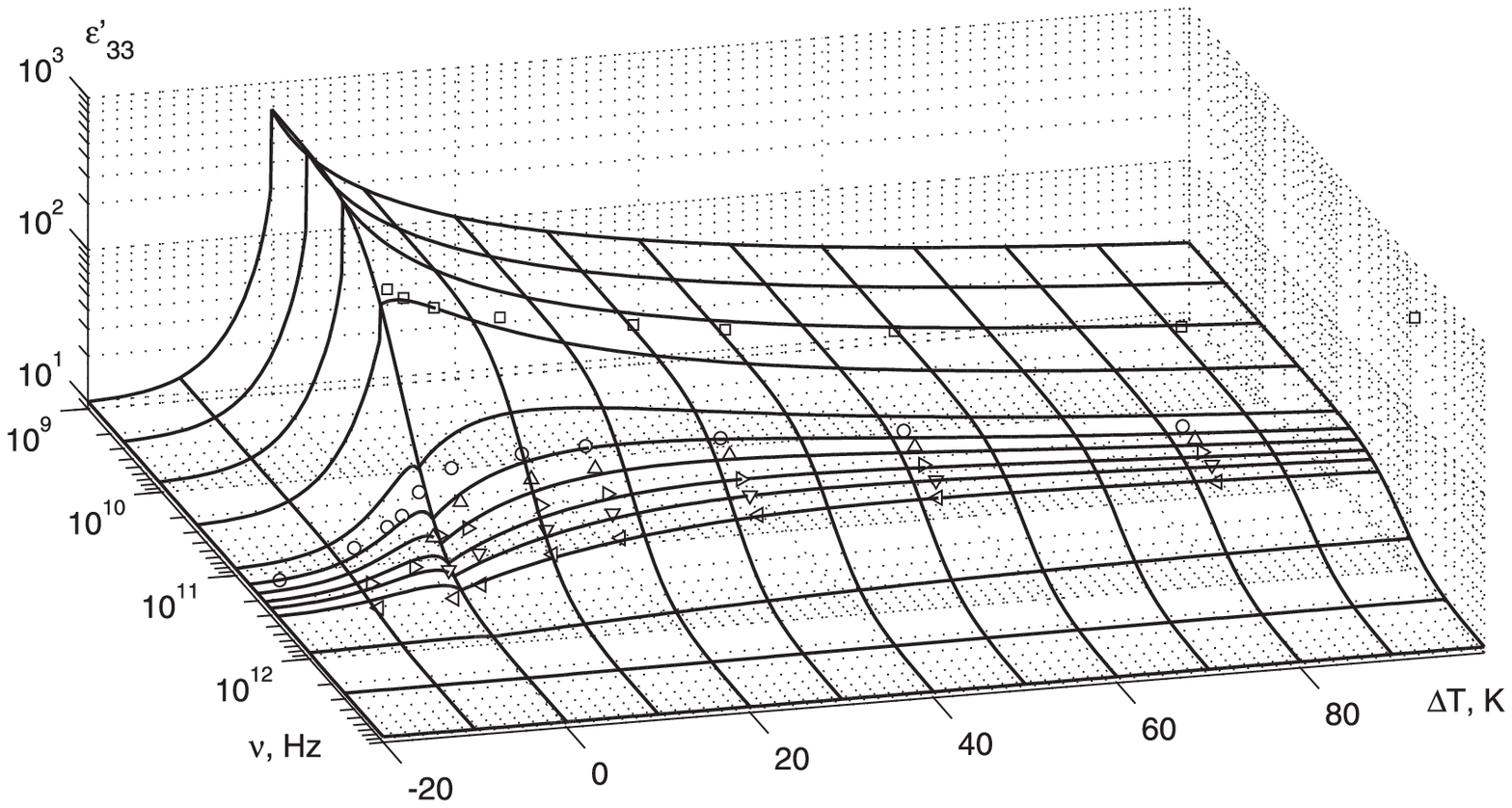}
\end{center}
\caption[]{The frequency-temperature dependence of $\varepsilon'_{33}$ in
KH$_{2}$AsO$_{4}$. \ra{\e{s0.eps}}~--~\cite{395x}; \ra{\e{o0.eps}},
\ra{\e{up0.eps}}, \ra{\e{ri0.eps}}, \ra{\e{do0.eps}},
\ra{\e{le0.eps}}~--~\cite{410x}. Symbols are experimental points; lines are the theoretical values.} \label{e33nuT_KDA}
\end{figure}
\begin{figure}[!h]
\begin{center}
 \includegraphics[width=0.7\textwidth]{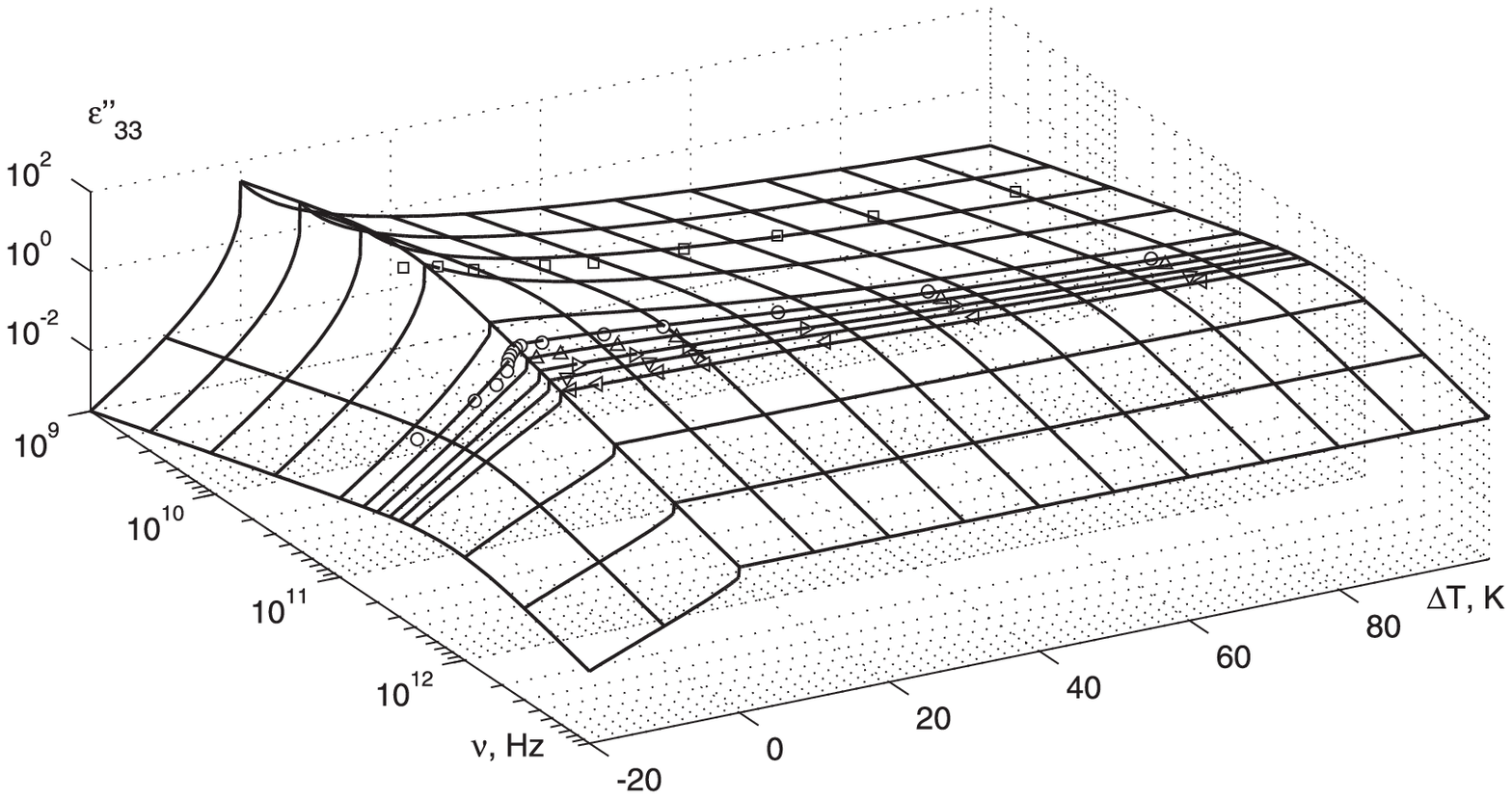}
\end{center}
\caption[]{The frequency-temperature dependence of $\varepsilon''_{33}$ in
KH$_{2}$AsO$_{4}$. \ra{\e{s0.eps}}~--~\cite{395x}; \ra{\e{o0.eps}},
\ra{\e{up0.eps}}, \ra{\e{ri0.eps}}, \ra{\e{do0.eps}},
\ra{\e{le0.eps}}~--~\cite{410x}.  Symbols are experimental points; lines are the theoretical values.} \label{e33nuTim_KDA}
\end{figure}

The most graphic illustration of the dispersion of the real and imaginary parts of the dielectric permittivity
 $\varepsilon^{*}_{33}(\omega,T)$ in  M(H$_{1-x}$D$_x)_2$XO$_4$ would be their frequency-temperature plots drawn in wide frequency and temperature ranges.
 Such plots for theoretical dependencies along with the experimental points are presented
 in figures~\ref{e33nuT_x093}, \ref{e33nuTim_x093} for
 K(H$_{0.07}$D$_{0.93})_2$PO$_4$, in figures~\ref{e33nuT_RDP}, \ref{e33nuTim_RDP} for RbH$_2$PO$_4$,
 and in figures~\ref{e33nuT_KDA}, \ref{e33nuTim_KDA} for KH$_2$AsO$_4$.

\begin{figure}[!h]
\begin{center}
 \includegraphics[scale=0.6]{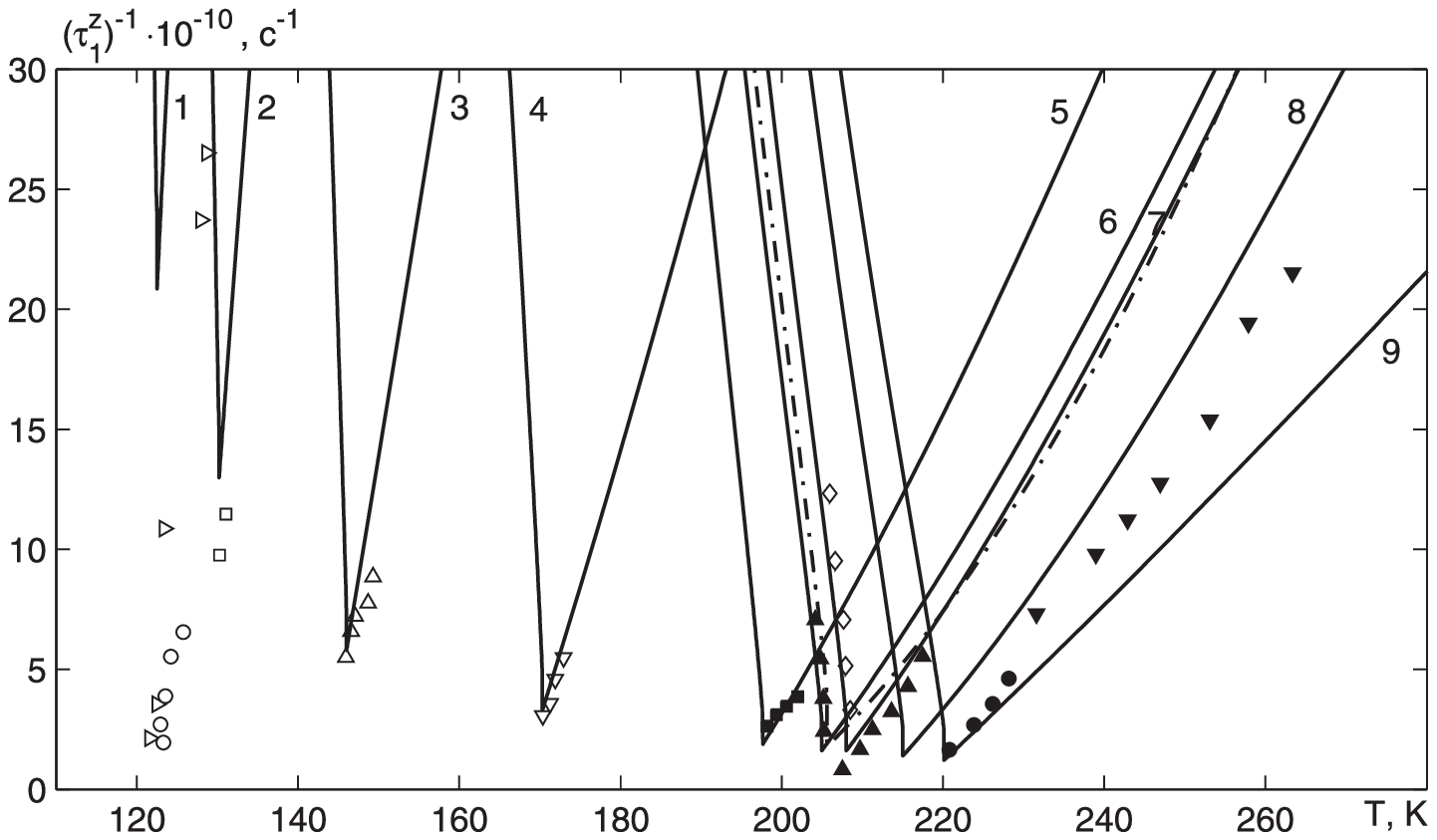}
\end{center}
\caption[]{The temperature dependence of the inverse polarization relaxation time at different $x$: 0.0~-- 1, \ra{\e{ri0.eps}} \cite{385x},
\ra{\e{le0.eps}} \cite{390x}, \ra{\e{o0.eps}} \cite{387x}; 0.07~-- 2,
\ra{\e{s0.eps}} \cite{391x}; 0.21~-- 3, \ra{\e{up0.eps}} \cite{391x};
0.43~-- 4, \ra{\e{do0.eps}} \cite{391x}; 0.72~-- 5,
\ra{\e{s1.eps}} \cite{391x}; 0.805~-- 6, \ra{\e{up1.eps}} \cite{378x};
0.84~-- 7, \ra{\e{d0.eps}} \cite{386x}; 0.93~-- 8,
\ra{\e{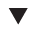}} \cite{390x}; 1.0~-- 9, \ra{\e{o1.eps}} \cite{394x}.
Symbols are experimental points; lines are the theoretical values.}
\label{taum1}
\end{figure}

Let us analyse the changes in the real and imaginary parts of
 $\varepsilon_{33}^{*}(\omega)$ in the  M(H$_{1-x}$D$_x)_2$XO$_4$ crystals
 at replacing H $\to$ D, K $\to$ Rb, and P $\to$ As. At $\Delta T =
 +0$~K, the dispersion frequency [i.e., the frequency of the maximum of  $\varepsilon''_{33}(\omega)$] is 33.2 in
 KH$_2$PO$_4$, 1.93 in KD$_2$PO$_4$,
20.5 in RbH$_2$PO$_4$, and 20.8 in KH$_2$AsO$_4$. The linewidth [i.e., the difference between frequencies of the maximum and half maximum of
 $\varepsilon''_{33}(\omega)$] is 12.0~GHz in  KH$_2$PO$_4$, 4.8 in KD$_2$PO$_4$,
 7.8 in RbH$_2$PO$_4$, and 7.2 in  KH$_2$AsO$_4$.

The temperature dependencies of the inverse relaxation time
 $(\tau_1^z)^{-1}$ in K(H$_{1-x}$D$_x)_2$PO$_4$ along with the values estimated from different experimental measurements are presented in figure~\ref{taum1}.
The calculated values of the relaxation times
$\tau_{2,3,4}^z$, in contrast to  $\tau_1^z$, are unlikely to depend on temperature and are much smaller than the value of $\tau_1^z$.
The theory provides a satisfactory agreement with the experiment for  temperature curves of the relaxation time. A certain
difference  between the relaxation times estimated from the dielectric permittivity and ultrasound measurements is due to the
contributions  into attenuation from the mechanisms irrelevant for the permittivity (e.g., scattering by admixtures).

\section{Conclusions}

Within the framework of the modified proton ordering model, taking into account a linear over the strain $\varepsilon _{6}$ contribution into the energy of the proton subsystem, and using the four-particle cluster approximation, we calculate the longitudinal dynamic characteristics of mechanically clamped crystals of the  KH$_{2}$PO$_{4}$ family. For the partially deuterated crystals M(H$_{1 -x}$D$_{x})_{2}$XO$_{4}$, these characteristics are obtained within the mean crystal approximation. The data  for $\varepsilon_{33}'(\nu,T)$ and $\varepsilon_{33}''(\nu,T)$ presented by  different groups of experimentalists are analyzed and systematized. At the proper choice of the theory parameters for the  M(H$_{1 -
x}$D$_{x})_{2}$XO$_{4}$ crystals, we obtain a good quantitative description of the available experimental data for $\varepsilon_{33}'(\nu,T)$ and $\varepsilon_{33}''(\nu,T)$.  For the first time, the dispersion of the longitudinal dynamic dielectric permittivity of clamped crystals of the  KH$_{2}$PO$_{4}$ family
is explored in wide temperature  and frequency ranges.
It should be noted that the effect of piezoelectric coupling on the dielectric characteristics of these crystals is essential. In the present paper, the observed temperature behavior of  $\varepsilon_{33}'(\nu,T)$ in the phase transition region at different frequencies has been appropriately described for the first time.

\newpage
\ukrainianpart

\title{Поздовжня релаксація механічно затиснутих кристалів типу KH$_2$PO$_4$}
\author{Р.Р. Левицький\refaddr{label1}, І.Р. Зачек\refaddr{label2}, А.С. Вдович\refaddr{label1}}

\addresses{
\addr{label1} Інститут фізики конденсованих систем НАН України вул.
Свєнціцького, 1, Львів, 79011, Україна, \addr{label2} Національний
університет ``Львівська політехніка'' вул. С. Бандери 12, 79013,
Львів, Україна }
%
%
%

\makeukrtitle

\begin{abstract}
\tolerance=3000%
У рамках модифікованої моделі протонного впорядкування сеґнетоактивних кристалів сім'ї КH$_{2}$PO$_{4}$ з врахуванням лінійного за деформацією внеску $\varepsilon_{6}$  в енергію протонної системи в наближенні чотиричастинкового кластера в межах динамічної моделі Глаубера отримано вираз для поздовжньої динамічної діелектричної проникності механічно затиснутого кристалу. При належному виборі параметрів теорії отримано добрий кількісний опис наявних експериментальних даних для цих кристалів.
\keywords сегнетоелектрики, кластерне наближення, діелектрична проникність, часи релаксації

\end{abstract}


\begin{thebibliography}{99}

\bibitem{145x} Levitsky~R.R., Zachek~I.R., Volkov.~A.A.,
Kozlov~G.V., Lebedev~S.P.  Preprint of the Bogolyubov Institute for
Theoretical Physics, ITP--80--13R, Kyiv, 1980 (in Russian).

\bibitem{58x} Yoshimitsu K., Matsubara T., Suppl. Progr. Theor.
Phys., 1968, \textbf{E68}, 109; \doi{10.1143/PTPS.E68.109}.

\bibitem{60x} Poplavko~Y.M., Physics of Dielectrics. Vyshcha Shkola, Kyiv, 1980 (in Russian).

\bibitem{16x} Vaks~V.G., Introduction into Microscopic Theory of Ferroelectrics. Moskow, 1973 (in Russian).

\bibitem{17x} Blinc R., Zeks B., Ferroelectrics and Antiferroelectrics.
Lattice dynamics. Moskow, 1975 (in Russian).

\bibitem{61x} Glauber J., J. Math. Phys., 1963, \textbf{4}, No.~2, 294; \doi{10.1063/1.1703954}.

\bibitem{342x} Levitsky R.R., Zachek I.R., Varanitsky V.I. Preprint of the Bogolyubov Institute for Theoretical Physics, ITP--79--11E, Kiev, 1979.

\bibitem{343x} Zachek~I.R., Levitsky~R.R.,  Teor. Mat. Fiz.,
1980, \textbf{43}, No.~ 1, 128 (in Russian) [Theor. Math. Phys., \textbf{43}, No.~1, 364; \doi{10.1007/BF01018473}].

\bibitem{344x} Levitsky~R.R., Zachek~I.R., Varanitsky~V.I., Ukr. J.~Phys., 1980, \textbf{25}, No.~12, 1961 (in Russian).

\bibitem{163x} Levitsky~R.R., Zachek~I.R., Mits~Ye.V.  Preprint of the Bogolyubov Institute for Theoretical Physics,
ITP--87--114R, Kyiv, 1987 (in Russian).

\bibitem{165x} Zachek~I.R., Mits~Ye.V., Levitsky~R.R.  Preprint of the Bogolyubov Institute for Theoretical Physics,
ITP--89--7R, Kyiv, 1987 (in Russian).

\bibitem{uni2009} Levitskii~R.R., Zachek~I.R., Vdovych~A.S., Sorokov~S.I.,  Condens. Matter Phys., 2009, \textbf{12}, No.~1, 75; \\ \doi{10.5488/CMP.12.1.75}.

\bibitem{48pok} Stasyuk~I.V., Levitskii~R.R., Korinevskii~N.A.,  Phys. Status Solidi B, 1979, \textbf{91}, No.~2, 541;\\ \doi{10.1002/pssb.2220910219}.

\bibitem{135x} Levitsky~R.R., Stasyuk~I.V., Korinevsky~H.A.,
Ferroelectrics, 1978, \textbf{21}, 481; \doi{10.1080/00150197808237303}.

\bibitem{137x} Korinevskii~N.A., Levitskii~R.R.,  Teor. Mat. Fiz., 1980, \textbf{42}, No.~3, 416 (in
Russian) [Theor. Math. Phys., 1980, \textbf{42}, No.~3, 274; \doi{10.1007/BF01018631}].


\bibitem{147pok} Yukhnovskii I.R., Levitskii~R.R., Sorokov~S.I., Derzhko O.V., Izv. AN SSSR, ser. fiz., 1991, \textbf{55}, No.~3, 481 (in Russian).

\bibitem{289pok} Levitskii~R.R., Sorokov~S.I., Baran~O.R., Condens. Matter Phys., 2000, \textbf{3}, No.~3, 515.

\bibitem{120pok} Levitskii~R.R., Sorokov~S.I.  Preprint of the Bogolyubov Institute for Theoretical Physics,
ITP--88--34R, Kyiv, 1988 (in Russian).


\bibitem{238pok} Levitskii~R.R., Sorokov~S.I., Moina A.P. Preprint of the Institute for
Condensed Matter Physics, ICMP--97--24U, Lviv, 1997 (in Ukrainian).



\bibitem{0311U1}
Yomosa~Sh., Nagamiya~T.,  Progr. Theor. Phys., 1949, \textbf{4},
No.~3, 263; \doi{10.1143/PTP.4.263}.

\bibitem{0311U2}
Slater~J.C.,  J. Chem. Phys., 1941, \textbf{9}, No.~1, 16; \doi{10.1063/1.1750821}.

\bibitem{0311U4}
Stasyuk~I.V., Biletskii~I.N.  Preprint of the Bogolyubov
Institute for Theoretical Physics, ITP--83--93R, Kyiv, 1983 (in
Russian).


\bibitem{0311U5}
Stasyuk~I.V., Biletskii~I.N., Styagar~O.N.,  Ukr. J. Phys., 1986,
\textbf{31}, No.~4, 567.

\bibitem{0311U6}
Stasyuk~I.V., Levitskii~R.R., Zachek~I.R., Moina~A.P.,  Phys. Rev.
B, 2000, \textbf{62}, No.~10, 6198; \\ \doi{10.1103/PhysRevB.62.6198}.

 \bibitem{Lis2003}
Levitskii~R.R., Lisnii~B.M.,  J. Phys. Stud., 2003, \textbf{7},
No.~4, 431 (in Ukrainian).

\bibitem{JPS1701}	Levitsky R.R., Zachek I.R., Vdovych A.S., Moina A.P., J. Phys. Stud., 2010, \textbf{14}, No.~1, 1701.

\bibitem{0311U8}
Levitskii~R.R., Lisnii~B.M.,  Phys. Status Solidi B, 2004,
\textbf{241}, No.~6, 1350.

\bibitem{0311U7}
Stasyuk~I.V., Levitskii~R.R., Moina~A.P., Lisnii~B.M.,
 Ferroelectrics, 2001, \textbf{254}, 213; \doi{10.1080/00150190108215002}.

\bibitem{lis2007}
Lisnii B.M.,  Levitskii R.R.,  Baran O.R., Phase Transitions, 2007, \textbf{80}, 25; \doi{10.1080/01411590701315591}.

\bibitem{Stasyuk2008} Stasyuk~I.V., Levitskii~R.R., Moina~A.P., Velychko~O.V.,  Ukr. J. Phys., 2008, \textbf{4}, 3 (in Ukrainian).

\bibitem{471x} Levitskii R.R., Lisnii B.M., J. Phys. Stud., 2002, \textbf{6}, No.~1, 91 (in Ukrainian).

\bibitem{cmp555} Levitsky R.R., Zachek I.R., Moina A.P., Vdovych A.S., Condens. Matter Phys., 2008, \textbf{11}, No.~3, 555.

\bibitem{533x} Stasyuk~I.V., Levitskii~R.R., Zachek~I.R., Vdovych~A.S., The SSS Physical Proceeding, 2011, \textbf{8}, 533.

\bibitem{icmp0608U} Levitskii~R.R., Zachek~I.R., Vdovych~A.S. Preprint of the Institute for
Condensed Matter Physics, ICMP--06--08U, Lviv, 2006 (in Ukrainian).

\bibitem{408x} Volkov~A.A., Kozlov~G.V., Lebedev~S.P.,
Velychko~I.A., Fiz. Tverd. Tela,  1979, \textbf{21}, No.~11, 3304
(in Russian).

\bibitem{395x} Kaminow I.P., Phys. Rev., 1965, \textbf{138}, A1539; \doi{10.1103/PhysRev.138.A1539}.

\bibitem{361x} Hill R.M., Ichiki S.K., Phys. Rev., 1963, \textbf{132}, No.~4, 1603; \doi{10.1103/PhysRev.132.1603}.

\bibitem{402x} Pereverzeva L.P., Poplavko Yu.M., Rez I.S., Kuznetsova L.I., Kristallografiya, 1976, \textbf{21}, No.~5, 981 (in Russian).

\bibitem{262x} Blinc R., Schmidt V.H.,
Ferroelectr. Lett. Sect., 1984, \textbf{1}, 119; \doi{10.1080/07315178408202409}.

\bibitem{403x} Pereverzeva L.P.,
Izv. AN SSSR, ser. fiz., 1971, \textbf{35}, No.~12, 2613 (in
Russian).

\bibitem{410x} Volkov A.A., Kozlov G.V., Lebedev S.P., Prokhorov A.M.,
Ferroelectrics, 1980, \textbf{25}, No.~1--4, 531; \doi{10.1080/00150198008207063}.

\bibitem{397x} Meriakri V.V., Ushatkin E.F. Investigation of inorganic materials by submillimeter spectroscopy metods. -- In: Physical methods of investigation of inorganic materials. Moscow, Nauka, 1981, p. 195--205 (in Russian).

\bibitem{399x} Gauss K.E., Happ H., Phys. Status Solidi B, 1976, \textbf{78}, No.~1, 133; \doi{10.1002/pssb.2220780111}.

\bibitem{352x} Skalyo J., Frazer B.C. Jr., Shirane G., Daniels W.B., J. Phys. Chem. Solids, 1969, \textbf{30}, No.~8, 2045; \doi{10.1016/0022-3697(69)90183-8}.

\bibitem{398x} Gauss K.E., Happ H., Rother G., Phys. Status Solidi B, 1975, \textbf{72}, No.~2, 623; \doi{10.1002/pssb.2220720220}.

\bibitem{396x} Meriakri V.V., Poplavko Yu.M., Ushatkin E.F., Zh. Tekh. Fiz., 1974, \textbf{44}, No.~5, 1111 (in Russian).

\bibitem{385x} Garland C.W., Novotny D.B., Phys. Rev., 1969, \textbf{177}, No.~2, 971; \doi{10.1103/PhysRev.177.971}.

\bibitem{390x} Vajda D., Acta Phys. Slov.,
1980, \textbf{30}, No.~1, 99.

\bibitem{387x} Litov E., Garland C.M., Phys. Rev. B, 1970, \textbf{2}, No.~11, 4597; \doi{10.1103/PhysRevB.2.4597}.

\bibitem{391x} Kasahara M., Tatsuzaki I., J. Phys. Soc. Jpn., 1981, \textbf{50}, No.~2, 551; \doi{10.1143/JPSJ.50.551}.

\bibitem{378x} Litov E., Uehling E.A., Phys. Rev. B, 1970, \textbf{1}, No.~9, 3713; \doi{10.1103/PhysRevB.1.3713}.

\bibitem{386x} Shimshoni M., Harnik E., Phys. Lett. A, 1970, \textbf{32}, No.~5, 321; \doi{10.1016/0375-9601(70)90526-8}.

\bibitem{394x} Reese R.L., Fritz J.J., Cummins H.Z.,
Phys. Rev. B, 1973, \textbf{7}, No.~9, 4165; \doi{10.1103/PhysRevB.7.4165}.


\end{thebibliography}
\end{document}